%% file: main.tex
\documentclass[sigconf]{acmart}

\usepackage{booktabs}   
\usepackage{multirow}   
\usepackage[normalem]{ulem} 
\useunder{\uline}{\ul}{}    
\usepackage{amsmath,amsfonts}
        
\usepackage{amssymb}    
\usepackage{algorithmic}
\usepackage{graphicx}
\usepackage{textcomp}
\usepackage{xcolor}
\usepackage{hyperref}  
\usepackage{svg}
\usepackage{subfigure}
\usepackage{float}      
\usepackage{bbm}    
\usepackage{enumitem}   

\usepackage[acronym]{glossaries}   

\usepackage{tabularx}

\usepackage{listings}
\definecolor{myred}{RGB}{178, 34, 34} 
\definecolor{mygreen}{RGB}{34,139,34}   
\definecolor{myred2}{RGB}{237, 211, 210} 
\definecolor{mygreen2}{RGB}{198, 232, 206} 
\definecolor{myblue2}{RGB}{218,232,252}

\definecolor{codegreen}{rgb}{0,0.6,0}
\definecolor{codegray}{rgb}{0.5,0.5,0.5}
\definecolor{codepink}{RGB}{252, 142, 172}
\definecolor{codepurple}{rgb}{0.58,0,0.82}
\definecolor{backcolour}{RGB}{245,245,245}

\lstdefinestyle{mystyle}{
    backgroundcolor=\color{backcolour},   
    commentstyle=\color{magenta},
    keywordstyle=\color{blue},
    numberstyle=\tiny\color{codegray},
    stringstyle=\color{codepurple},
    basicstyle=\fontfamily{\ttdefault}\footnotesize,
    breakatwhitespace=false,         
    breaklines=true, 
    breakindent=0pt,     
    keepspaces=true,    
    frame=single, 
    numbersep=5pt,                  
    showspaces=false,                
    showstringspaces=false,
    showtabs=false,                  
    tabsize=2,
    classoffset=1, 
    keywordstyle=\color{violet},
    classoffset=0,
}
\lstdefinelanguage{JavaScript}{
  keywords={typeof, new, true, false, catch, function, return, null, catch, switch, var, if, in, while, do, else, case, break},
  keywordstyle=\color{blue}\bfseries,
  ndkeywords={class, export, boolean, throw, implements, import, this},
  ndkeywordstyle=\color{darkgray}\bfseries,
  identifierstyle=\color{black},
  sensitive=false,
  comment=[l]{//},
  morecomment=[s]{/*}{*/},
  commentstyle=\color{purple}\ttfamily,
  stringstyle=\color{red}\ttfamily,
  morestring=[b]',
  morestring=[b]"
}

\definecolor{correctcolor}{RGB}{34, 139, 34} 
\definecolor{wrongcolor}{RGB}{220, 20, 60}   
\definecolor{partialcolor}{RGB}{255, 140, 0} 

\lstdefinestyle{mongostyle}{
    basicstyle=\ttfamily\footnotesize,
    breaklines=true,
    postbreak=\mbox{\textcolor{red}{$\hookrightarrow$}\space},
}
\newcommand{\compactcode}[1]{\lstinline[style=mongostyle]|#1|}
\newcommand{\correct}[1]{\textcolor{correctcolor}{\textbf{#1}}}
\newcommand{\wrong}[1]{\textcolor{wrongcolor}{\textbf{#1}}}
\newcommand{\partialcorrect}[1]{\textcolor{partialcolor}{\textbf{#1}}}

\AtBeginDocument{
  }

\newcommand{\DatasetName}{CoNoSQL}
\newcommand{\MethodName}{Stage-MCTS}

\setcopyright{none}
\renewcommand\footnotetextcopyrightpermission[1]{}
\begin{document}
\title{Monte Carlo Tree Search with Reasoning Path Refinement for Small Language Models in Conversational Text-to-NoSQL}

\author{Xubang Xiong$^{1}$, Raymond Chi-Wing Wong$^{1}$, Yuanfeng Song}
\affiliation{
\institution{$^{1}$The Hong Kong University of Science and Technology, Hong Kong, China}
\country{}}

\renewcommand{\authors}{Xubang Xiong, Raymond Chi-Wing Wong, Yuanfeng Song}

\input{sections/abstract}

\maketitle

\input{sections/introduction}

\input{sections/preliminaries}

\input{sections/method}

\input{sections/experiment}

\input{sections/related_work_short}

\input{sections/conclusion}

\bibliographystyle{ACM-Reference-Format}
\bibliography{reference/conosql}

\input{sections/appendix}

\end{document}

%% file: sections/abstract.tex
\begin{abstract}
NoSQL databases have been widely adopted in big data analytics, geospatial applications, and healthcare services, due to their flexibility and scalability. However, querying NoSQL databases requires specialized technical expertise, creating a high barrier for users. While recent studies have explored text-to-NoSQL problem, they primarily focus on single-turn interactions, ignoring the conversational nature of real-world queries. To bridge this gap, we introduce the Conversational Text-to-NoSQL task, which generates NoSQL queries given a natural language question, a NoSQL database, and the dialogue history. To address this task, we propose \MethodName, a framework that endows small language models (SLMs) with NoSQL-specific reasoning capabilities by formulating query generation as a search problem. The framework employs Monte Carlo Tree Search (MCTS) guided by a rule-based reward to produce stepwise reasoning data, followed by progressive supervised fine-tuning (SFT) and self-training strategies. We further construct \DatasetName, a cross-domain dataset with over 2,000 dialogues and 150 databases, to support evaluation. Experiments demonstrate that our approach outperforms state-of-the-art large reasoning models, {improving execution value match (EVM) accuracy by up to $7.93\%$.}
\end{abstract}

%% file: sections/introduction.tex
\section{Introduction}  \label{sec:introduction}

Not only SQL (NoSQL) database is an advanced data storage and management system that stores data in a non-tabular format, designed to store data without a pre-defined structure. It attaches great importance to big data analytics~\cite{MH+2013,KSM+2017}, geospatial applications~\cite{BDG+2017,GO2020} and healthcare services~\cite{TBT+2019,YLH+2013,SM2024}, due to its high performance in processing large-scale data flexibly.

Despite the numerous advantages of NoSQL databases, users are required to learn specialized knowledge to write query statements and manage data effectively, which presents a high technical barrier. To address this challenge, querying the NoSQL database with natural language interfaces is proposed, which facilitates more intuitive interactions between the users and the NoSQL databases. The existing research~\cite{LSQ+2025,QSL+2025} explores the text-to-NoSQL problem. Given a natural language question and the NoSQL database information, the system is to predict NoSQL query. However, the study assumes that only one natural language question can be mapped into the final NoSQL query to meet the requirements of the user, within the context of a single-turn interaction scenario. 

\begin{figure}
    \centering
    \includegraphics[width=0.95\linewidth]{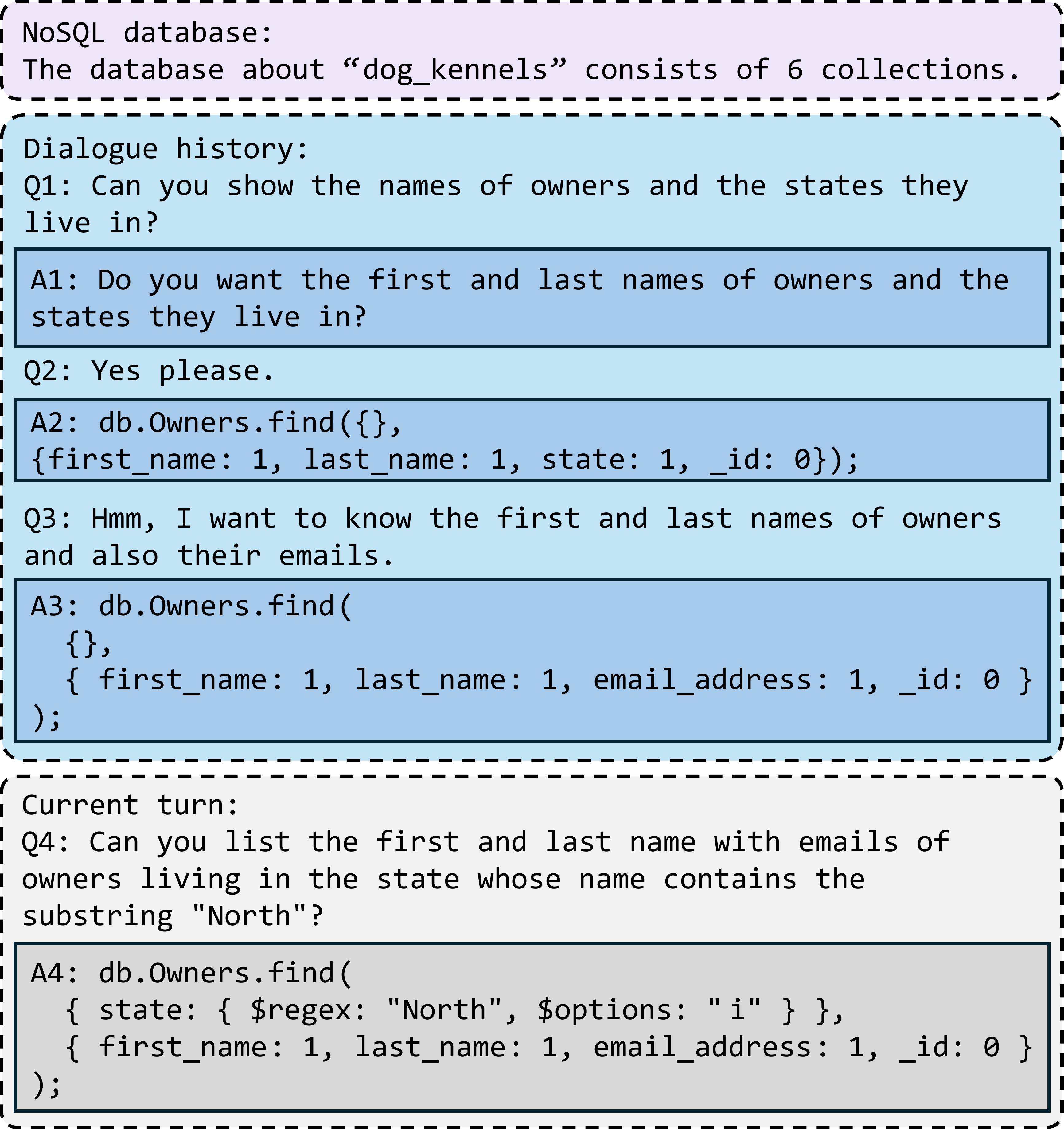}
    \vspace{-5pt}
    \caption{A dialogue from the \DatasetName~dataset. MongoDB is utilized in this scenario as a representative of NoSQL database. $Q_{i}$ represents the user question in turn $i$ and $A_{i}$ refers to the response (i.e., usually a NoSQL query) from the system.}
    \label{fig:introduction_example}
    \vspace{-15pt}
\end{figure}

In practice, real-world scenarios rarely involve isolated queries. Instead, most practical interactions consist of multiple continuous and multi-turn exchanges. It is essential for the effective human-AI communication systems to consider the conversational context, interpret previous exchanges with nuanced understanding and iteratively refine the objectives. Consequently, the single-turn interaction paradigm presents a significant limitation, restricting the full exploitation of the advanced capabilities inherent in the natural language interfaces of NoSQL databases. Motivated by this, we propose \textbf{Conversational Text-to-NoSQL} problem. Given a natural language question, the corresponding NoSQL database information and the dialogue history, the system is to generate the NoSQL query to satisfy the requirements of the user.

As shown in Figure~\ref{fig:introduction_example}, given the current-turn user question, the system generates a NoSQL query using the information from the NoSQL database about "dog\_kennels" and the dialogue history of all context-relevant exchanges. It indicates that this paper studies how to enhance the NoSQL database with natural language interface to generate a NoSQL query that is tailored to extract the specific information requested by the user, taking into account both the database schema and the ongoing conversation context.

Unlike traditional SQL queries, NoSQL queries rely on an aggregation pipeline for data processing. Furthermore, NoSQL databases support a dynamic schema, meaning that documents in the same collection can have different structures. These unique characteristics render the conversational text-to-NoSQL problem notably challenging. To address this problem while balancing efficiency and effectiveness, a straightforward approach is to deploy small language models (SLMs) for NoSQL query generation under the Chain-of-Thought (CoT)~\cite{WWS+2022,KGR+2022} framework. CoT reasoning has been shown to drive substantial performance gains for language models across diverse tasks, including mathematical reasoning~\cite{GSY+2025,YJS+2024} and scientific problem-solving~\cite{SHZ+2024,WHL+2024}. However, preliminary experiments presented in Table~\ref{tab:preliminary_exp_NoSQL_reasoning} of the appendix reveal that SLMs employing CoT even underperform the methods without CoT reasoning. This phenomenon can be attributed to the inherent lack of NoSQL-specific reasoning capabilities in SLMs.

Motivated by these observations, we propose \emph{\MethodName}, a framework designed to endow SLMs with NoSQL-specific reasoning capabilities inherently. Specifically, we formulate the conversational Text-to-NoSQL problem as a search problem. The SLM first generates executable NoSQL stages incrementally, where each stage is paired with a corresponding natural language comment, and then assembles these pre-generated stages to produce the final query.
To acquire extensive high-quality training data, an SLM augmented with Monte Carlo Tree Search (MCTS) is leveraged to generate multiple reasoning paths, guided by a rule-based reward model. MCTS decomposes complex conversational Text-to-NoSQL tasks into more manageable single-step generation subtasks. Furthermore, its inherent stepwise generation mechanism naturally facilitates the derivation of step-level training data. These generated paths are further filtered by the proposed reward-based sampling and path refinement strategies to ensure data quality.
After constructing the training dataset, the SLM undergoes progressive supervised fine-tuning (SFT), namely three-phase SFT. It aims at enabling the SLM to learn knowledge incrementally from easy to complex. Given the inherent limitations of SLMs in specialized reasoning, a multi-round self-training pipeline is adopted to iteratively generate high-quality data and enrich the training set with diverse reasoning paths. After multiple rounds of self-training, the updated SLM is employed to predict NoSQL queries via an MCTS-based test-time scaling method, where query generation is guided by a broadly recognized self-supervised reward model~\cite{LZF+2025,PLS+2025}.

To validate the effectiveness of the proposed framework, we construct a cross-domain conversational text-to-NoSQL dataset for dedicated experimental evaluation, namely \DatasetName. This dataset encompasses over 2,000 dialogues, more than 9,000 turn-based exchanges, and over 150 cross-domain databases. By spanning a wide range of real-world application scenarios, \DatasetName~effectively ensure the generalizability and reliability of the experimental results.

The key contributions of this work are summarized as follows.
\begin{itemize}
    \item To the best of our knowledge, this is the first work to propose the conversational text-to-NoSQL task and to formulate NoSQL query generation as a search problem.

    \item The proposed \MethodName~framework is proven to be effective in endowing SLMs with tailored NoSQL-specific reasoning capabilities that are lacking in existing models.
    
    \item A novel dataset, namely \DatasetName, is built to validate the efficacy of \MethodName. This dataset leverages the collaborative effort of LLMs and human evaluation, thereby enhancing the efficiency of dataset curation while ensuring the production of high-quality and reliable data.
    
    \item Extensive simulations validated the viability of {\MethodName} and the analysis of performance and ablation study give guidance to use the framework. Notably, \MethodName~surpasses the advanced reasoning models such as DeepSeek-R1 and GPT-4o in Execution Value Match (EVM) accuracy, {with the maximum improvement reaching $7.93\%$.}
\end{itemize}

The rest of the paper is organized as follows. Section \ref{sec:preliminaries} presents the necessary preliminaries. Section \ref{sec:method} elaborates on the design and analysis of the proposed framework, along with the construction of the validation dataset. Section \ref{sec:experiment} demonstrates the performance of the proposed framework. Finally, Section \ref{sec:related_work} reviews relevant research, and Section \ref{sec:conclusion} summarizes the key findings of this paper.

%% file: sections/preliminaries.tex
\section{Preliminaries}     \label{sec:preliminaries}
\subsection{Task} \label{subsec:task}

The task of conversational text-to-NoSQL is formally defined as follows. At each interaction turn $i$, given the current user question $q_i$, the NoSQL database information (e.g., schema) $\mathcal{D}$ and the dialogue history $\mathcal{H}_i=\{(q_1,y_1), ... (q_{i-1},y_{i-1})\}$, the model generates the NoSQL query $y_i$ to meet the requirement of the user.

\subsection{NoSQL Query Languages} \label{subsec:nosql_queries}
NoSQL and SQL databases differ fundamentally in their query paradigms, which stems from their underlying data models and design philosophies. 

The core variance between the two query languages lies in their query execution stages. Taking MongoDB as an example, it leverages an aggregation pipeline, illustrated in Figure~\ref{mongodb_query}. This pipeline consists of multiple stages. Each stage processes the incoming documents and forwards the transformed results to the next stage.
In this example, the $\$group$ operator acts on the data already filtered by the $\$match$ operator, rather than directly operating on the raw data stored in the "orders" collection.

\begin{figure}[htbp]  
\centering
\begin{lstlisting}[breaklines=true]
 db.orders.aggregate([
    { $match: { orderDate: { $gte: ISODate("2023-01-01"), $lt: ISODate("2024-01-01") } } },
    { $group: { _id: "$customerId", totalSales: { $sum: "$amount" } } },
    { $sort: { totalSales: -1 } }
  ])
\end{lstlisting}
\vspace{-10pt}
\caption{NoSQL query example.}
\vspace{-10pt}
\label{mongodb_query}
\end{figure}

%% file: sections/method.tex
\section{Methodology} \label{sec:method}
\subsection{Overview}
\begin{figure*}[htbp]
    \centering
    \includegraphics[width=0.95\textwidth]{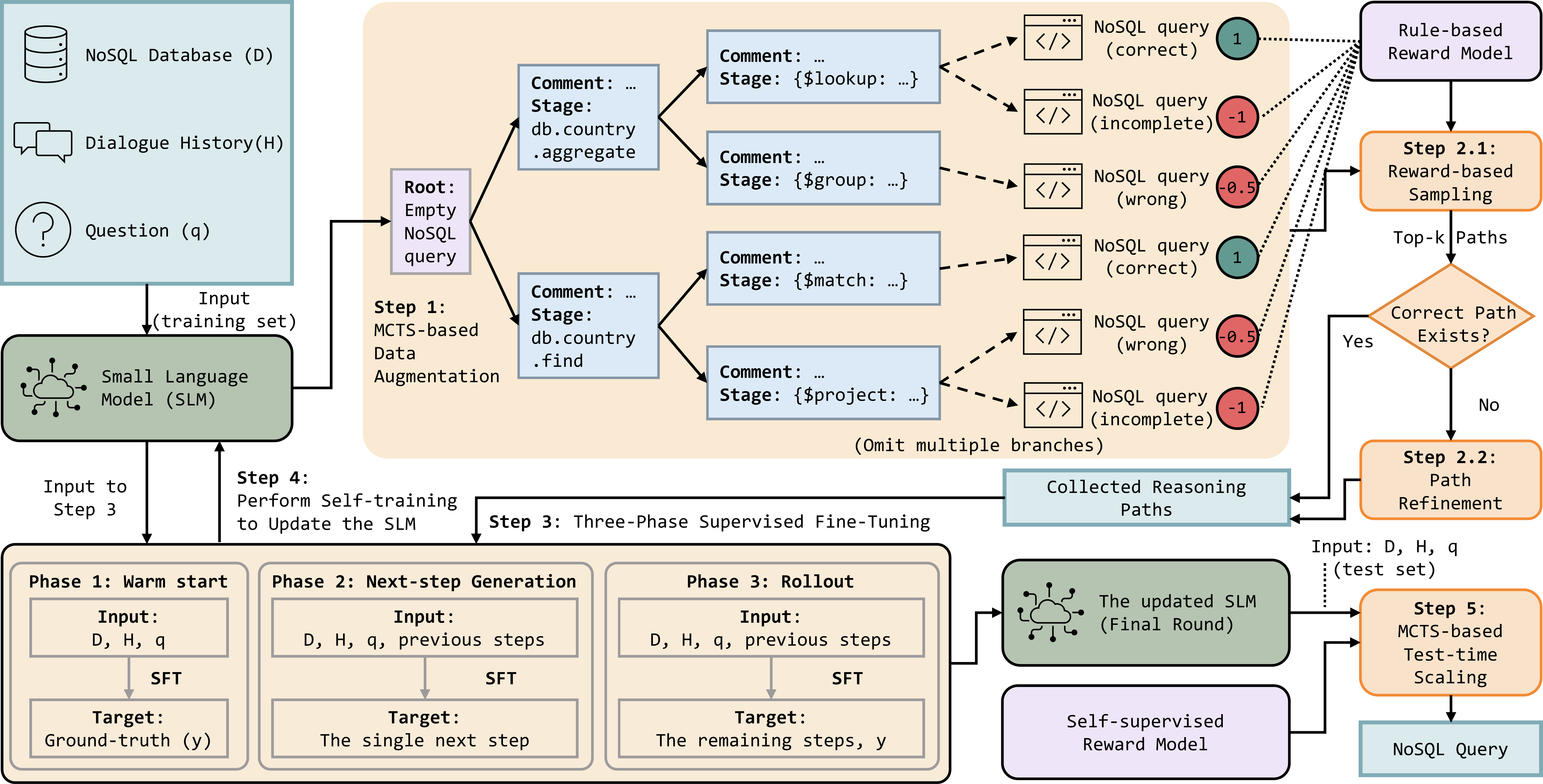}
    \vspace{-5pt}
    \caption{A small language model (SLM) is employed to generate multiple reasoning paths using a rule-based reward model. From these paths, the top-k correct paths are selected via reward-based sampling. If no correct path is yielded in a sampling step, the incorrect paths are refined with the ground-truth. The collected reasoning paths are subsequently used to iteratively train the SLM. Finally, the self-trained SLM is applied to predict the NoSQL query with test-time scaling based on MCTS.}
    \vspace{-5pt}
    \label{fig:method_overview}
\end{figure*}

An SLM is leveraged to explore the generation of higher-quality training data. To enhance NoSQL reasoning capabilities, a stage-augmented CoT data augmentation method is proposed, presented in Section~\ref{subsec:SoT}. As illustrated in Figure~\ref{fig:method_overview}, an SLM integrated with MCTS generates multiple reasoning paths under the guidance of a rule-based reward model. From these paths, the top-$k$ correct ones are selected through reward-based sampling. If no valid reasoning path is corrected in a given sampling step, the incorrect paths are refined using the ground-truth. These collected reasoning paths are subsequently incorporated into the training set, which is used to conduct three-phase fine-tuning, as illustrated in Section~\ref{subsec:three_phase_sft}. The fine-tuned model then serves to generate higher-quality training data in the next iteration, shown in Section~\ref{subsec:self_training_pipeline}. Finally, Section~\ref{subsec:mcts_based_tts} describes how the self-improved SLM is applied to predict NoSQL queries with MCTS-based test-time scaling. To validate the effectiveness of the proposed method, we construct CoNoSQL, a cross-domain conversational text-to-NoSQL dataset, for evaluation.

\subsection{Stage-augmented CoT Generation} \label{subsec:SoT}
\subsubsection{Framing Text-to-NoSQL as a Search Problem}
The conversational Text-to-NoSQL task could be defined as a search problem over a vast space of potential stages. Different from SQL queries, NoSQL queries (e.g., MongoDB queries) involve multiple stages embedded within methods including aggregate, find, distinct, and count, as elaborated in Section~\ref{subsec:nosql_queries}.

\begin{figure}
    \centering
    \includegraphics[width=1.0\linewidth]{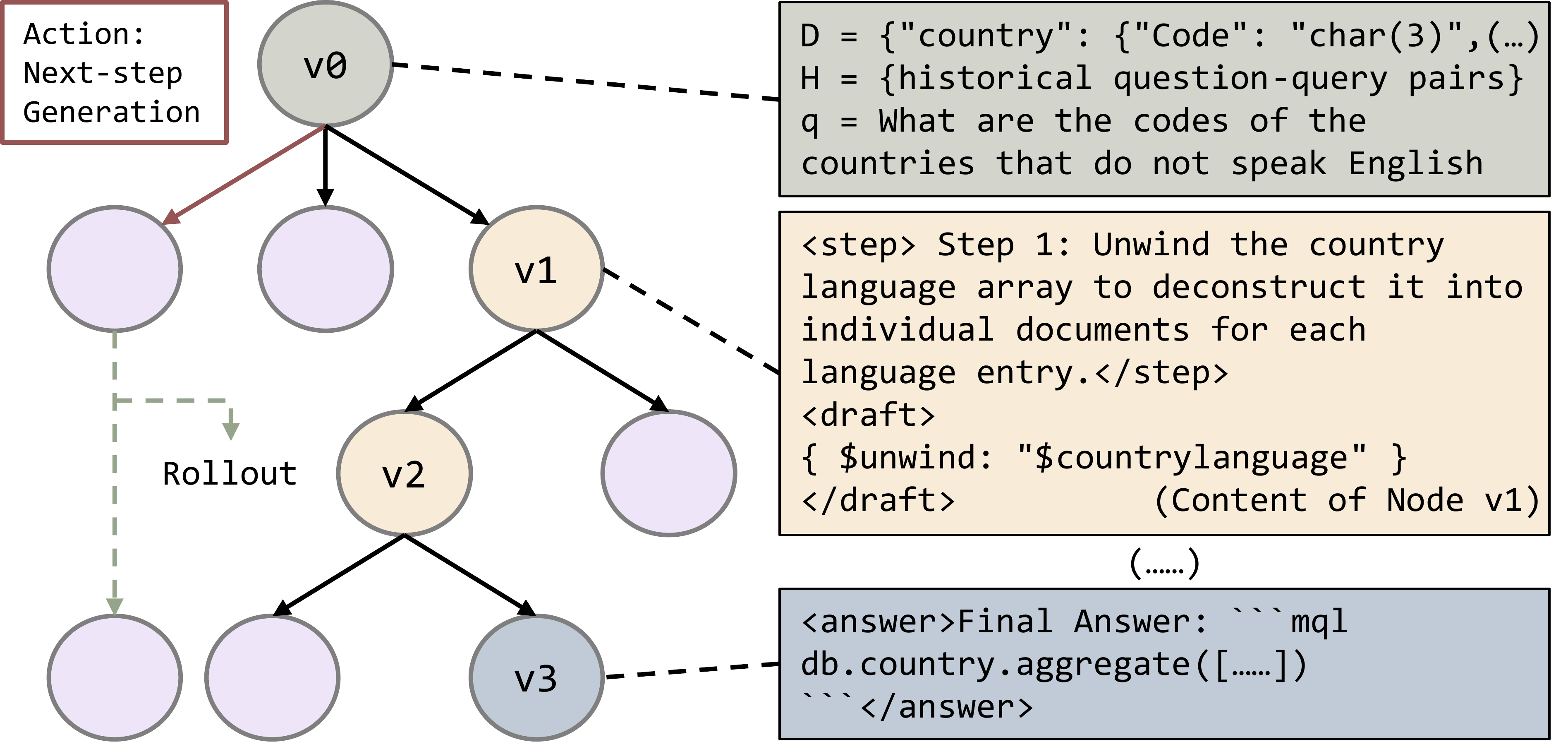}
    \vspace{-5pt}
    \caption{Framing Text-to-NoSQL as a Search Problem}
    \label{fig:mcts_search_problem}
    \vspace{-5pt}
\end{figure}

Thus, the search space $\mathcal{S}$ consists of all possible combinations of these stages, given a NoSQL database $\mathcal{D}$, the dialogue history $\mathcal{H}_i$ and the current question $q_i$. This space is represented as a tree structure $\boldsymbol{\Phi}=(V,E)$, where:

\noindent \textbf{Nodes ($V$).}
Each node $v \in V$ denotes a partial reasoning state corresponding to a specific step in the query construction process, shown in Figure~\ref{fig:mcts_search_problem}. Specifically, the \emph{root node} $v_0$ denotes the initial empty query and encapsulates input information including the NoSQL database $\mathcal{D}$, the dialogue history $\mathcal{H}_i$ and the current question $q_i$. An \emph{intermediate node}, e.g., $v_1$, stores a comment linked to the generated stage that contributes to the final NoSQL query construction. A \emph{leaf node} $v_d$, exemplified here by $v_3$, denotes either a fully constructed NoSQL query or a single generated stage.

\noindent \textbf{Edges ($E$).}
Each edge $e \in E$ denotes an action in the query construction process, such as generating a stage or constructing the final NoSQL query. These actions model transitions between intermediate query states within the search tree.

In the search tree, any root-to-leaf path that ends at a terminal node forms a trajectory $\boldsymbol{t}=v_0\oplus v_1\oplus\dots\oplus v_d$, where $\oplus$ denotes the concatenation of the actions associated with edges along the path.

\subsubsection{Stage-augmented Chain-of-Thoughts} 
Prior approaches primarily generate natural language CoT~\cite{WWS+2022,KGR+2022}. However, in the conversational Text-to-NoSQL task, reasoning steps need to be tailored to the unique characteristics of NoSQL-specific reasoning, thereby helping the model better grasp the task requirements. To address this, stage-augmented CoT is proposed. As shown in Figure~\ref{fig:mcts_search_problem}, each reasoning step corresponds to an executable NoSQL stage paired with a natural language comment. As an integral component of the final query, this design effectively enhances the interpretability.

\subsubsection{Improved MCTS for Data Augmentation} 
This study focuses on training an SLM with reasoning capabilities for the conversational Text-to-NoSQL task. Specifically, the SLM first generates executable NoSQL stages step by step, each paired with a corresponding natural language comment. It then assembles these pre-generated stages to output the final query. To address the limitation of requiring extensive training data with detailed reasoning steps, we propose leveraging MCTS. First, it decomposes complex conversational Text-to-NoSQL tasks into more manageable single-step generation subtasks, effectively lowering the overall complexity of the problem. Furthermore, building on our proposed framework, where Text-to-NoSQL is formulated as a search problem, the stepwise generation inherent to MCTS naturally facilitates the derivation of step-level training data. This dual advantage not only simplifies the core generation challenge but also provides structured, granular data to enhance model training and generalization. The details of the improved MCTS algorithm are as follows.

\noindent \textbf{Node Selection.}
The well-known Upper Confidence bounds for Trees (UCT)~\cite{KS2006} is employed to select the best node among the candidates, i.e., the one with the maximum $\operatorname{UCT}(v)$ value. This selection process is mathematically denoted as:
\begin{equation}
\operatorname{UCT}(v)=Q(v)+c \sqrt{\frac{\ln N_{\text {parent }}(v)}{N(v)}} ; \quad \text { where } \quad Q(v)=\frac{q(v)}{N(v)}.
\end{equation}
In this process, $N(v)$ denotes the number of visits to node $v$, and $N_{\text {parent }}(v)$ denotes the number of visits to $v$'s parent node. The predicted reward $q(v)$ is the output of the reward model, discussed in the following section and updated via backpropagation. $c$ is a constant that balances exploitation and exploration.

\noindent \textbf{Stage Expansion.}
Given the selected node $v_{select}$, the SLM generates $m$ child nodes, where $m$ is a positive integer hyper-parameter. Each child node contains a single stage and its corresponding reasoning, derived from the previous steps stored in $v_{select}$.

\noindent \textbf{Threshold-based Rollout.}
A complete rollout process involves iterative node selection and stage expansion until a termination node is reached. However, a complex NoSQL query may involve multiple stages, and generating these stages during training data sampling can be time-consuming. To balance the efficiency and effectiveness of MCTS, we propose a threshold-based rollout step. Specifically, when the depth of the selected node $v_{select}$ reaches a pre-defined positive integer threshold $max\_depth$, the policy SLM generates all remaining stages and the final NoSQL query in a single response. All newly expanded child nodes are persistently retained in the tree structure throughout the rollout process.

\noindent \textbf{Backpropagation.} At the termination node, backpropagation commences with the evaluation of the \emph {Execution Value Match (EVM)} metric, based on the predicted and ground-truth NoSQL queries. EVM assesses whether the values in the predicted query’s execution results match those of the ground-truth query, since it eliminates interference from custom field names and focuses solely on unique values. The detailed definition of EVM is provided in Section~\ref{sec:appendix_exp_metrics} in the appendix. Consider the trajectory $\boldsymbol{t}=v_0\oplus v_1\oplus\dots\oplus v_d$ from the root node $v_0$ to the termination node $v_d$. For each node $v$ on this trajectory, the predicted reward $q(v)$ and visit counts $N(v)$ are updated as: $q(v)=q(v)+r, N(v)=N(v)+1$. The reward $r$ is defined as $1.0$ if EVM is true, $-0.5$ if EVM is false, and $-1.0$ for the unexecutable predicted NoSQL query. This widely adopted rule-based reward model~\cite{TDG+2025,GYZ+2025} is both accurate and efficient.

\subsubsection{Reward-based Sampling and Path Refinement}
After $N_{rollout}$ rollouts, multiple complete reasoning trajectories are collected, forming a set of candidate NoSQL query reasoning trajectories $\mathcal{T}=\{\boldsymbol{t}_1,\dots,\boldsymbol{t}_n\}$. Here, $N_{rollout}$ is a user-defined hyper-parameter. Among these candidate trajectories, we filter and retain the top-$k$ correct paths that yield a reward of $1.0$. In cases where no valid reasoning path is derived for a specific generation round, the erroneous paths are refined based on the ground-truth query, leveraging the SLM. Specifically, given a NoSQL database $\mathcal{D}$, the dialogue history $\mathcal{H}_i$, the current question $q_i$ and the ground-truth query $y_i$, the SLM generates all reasoning paths in a single response. The prompt for path refinement via SLM is presented in Figure~\ref{prompt:path_refinement} in the appendix.

\begin{figure*}[tp]
    \centering
    \includegraphics[width=0.85\textwidth]{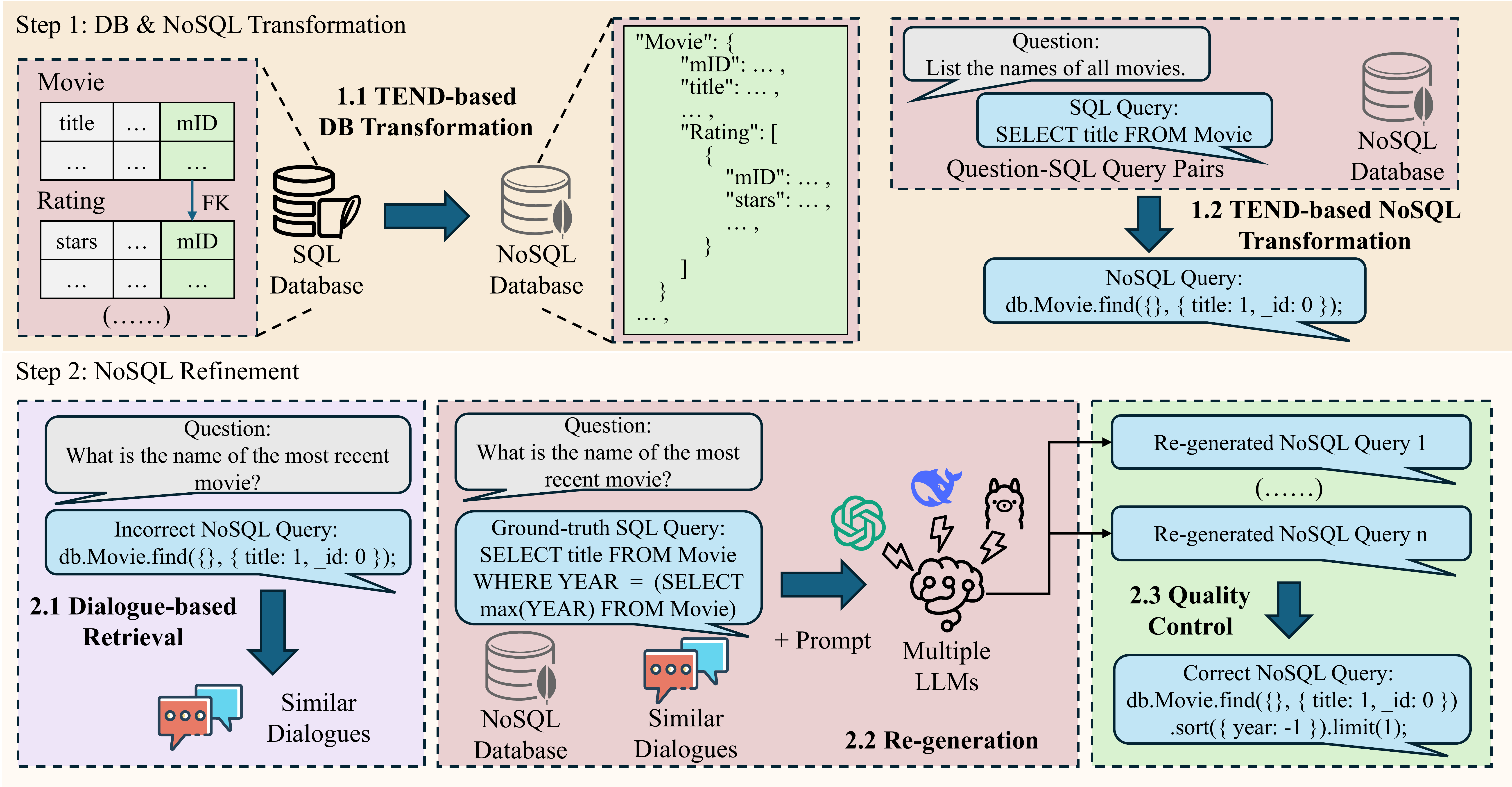}
    \vspace{-10pt}
    \caption{The overview of the pipeline of dataset construction. Data from relational databases is converted to NoSQL-compatible data. Incorrectly transformed data is refined using a dialogue-based RAG approach. The quality control measures, including manual review, are implemented to ensure data quality.}
    \label{fig:dataset_pipeline}
\end{figure*}

\subsection{Three-Phase Supervised Fine-Tuning} \label{subsec:three_phase_sft}
As illustrated in Figure~\ref{fig:method_overview}, the SLM undergoes a three-phase SFT process, namely warm start, next-step generation, and rollout.

In Phase 1 (warm start), the input instruction is restricted to the information encapsulated in the root node, which comprises a NoSQL database $\mathcal{D}$, the dialogue history $\mathcal{H}_i$ and the current question $q_i$. The corresponding target response is the ground-truth NoSQL query $y_i$. This phase is designed to familiarize the SLM with the syntax and semantics of NoSQL queries.
In Phase 2 (next-step generation), the input instruction is augmented with both the root node information and all previously generated steps. The target response for this phase is a single next step, consisting of an executable NoSQL stage paired with a natural language comment. Its objective is to enhance the ability of the SLM to follow the instructional guidance for step-wise generation.
In Phase 3 (rollout), the input instruction is augmented with both the root node information and the sequence of previously generated steps, with the objective of endowing the SLM with the capability to generate all remaining steps together with the final NoSQL query. This also serves as the target response of this phase.
The core objective of SFT is to minimize the negative log-likelihood loss ($\mathcal{L}_{\mathrm{SFT}}$) over the training set, formulated as follows:
\begin{equation}
\mathcal{L}_{\mathrm{SFT}}=-\frac{1}{L} \sum_{t=1}^L \log P\left(y_t \mid x, y_{<t} ; \theta\right),
\end{equation}
where $L$ denotes the token length of the target response $y$, $x$ represents the input instruction and $\theta$ denotes the learnable parameters of the language model.

\subsection{Self-training Pipeline} \label{subsec:self_training_pipeline}
Given the limited reasoning capabilities of SLMs, we adopt a multi-round self-training framework to iteratively generate high-quality data and enrich the training set with diverse reasoning paths. Specifically, the self-training process consists of two key steps. 
The first is stage-augmented CoT generation, where the SLM is leveraged to sample a variety of reasoning paths. These newly sampled paths are incorporated into the original training set, with the updated training set restricted to the data generated in the two most recent rounds for lower training complexity and computational costs. Duplicate reasoning data is also removed.
The second is three-phase SFT, in which the refined training set is used to adaptively fine-tune the SLM. The updated SLM is then deployed in the subsequent round to produce training data of even higher quality.
Full details can be found in Section~\ref{subsubsec:appendix_self-training} in the appendix.

\subsection{MCTS-based Test-time Scaling} \label{subsec:mcts_based_tts}
After undergoing multiple rounds of self-training, the updated SLM is employed to predict the NoSQL queries. To further enhance the reliability of the final answer, the MCTS algorithm from Section~\ref{subsec:SoT} is leveraged to scale up test-time computation. However, the ground-truth queries are unavailable during the inference phase. To implement the reward-guided search strategy, a self-supervised reward model is adopted, which has been proven effective in numerous studies~\cite{LZF+2025,PLS+2025}. This reward is calculated as follows:
\begin{equation}
R(y, q_i, \mathcal{H}_i, \mathcal{D})=\frac{1}{N} \sum_{j=1}^N \mathbbm{1}\left[\operatorname{Execute}(y, \mathcal{D})=\operatorname{Execute}\left(y_j, \mathcal{D}\right)\right]
\end{equation}
where $\mathbbm{1}$ denotes the indicator function, $\operatorname{Execute}$ is the operator that executes the query on the given database, $y_j$ are sampled NoSQL queries, and $N$ is the number of samples. This reward model prioritizes NoSQL queries with consistently stable execution outcomes (i.e., EVM), empowering MCTS to focus on reliable reasoning trajectories without the need for annotated data.
Regarding the selection model, a greedy selection strategy is adopted to identify the optimal NoSQL query. Specifically, after the MCTS algorithm generates a diverse set of candidate reasoning paths, the NoSQL query corresponding to the path that yields the highest reward score is selected.

\input{sections/verification}

%% file: sections/verification.tex
\subsection{Framework Verification}

\subsubsection{Dataset Construction}
Figure~\ref{fig:dataset_pipeline} illustrates the overview of the dataset construction. Initially, the method introduced in the study of the dataset, TEND~\cite{LSQ+2025}, was employed to transform the relational databases into the NoSQL databases and translate the SQL queries to the NoSQL queries. Subsequently, a retrieval augmented generation (RAG)~\cite{LPP+2020, GXG+2023, ZZY+2024} supported by multiple large language models (LLMs) is proposed to refine the transformed NoSQL queries. Eventually, the quality control strategies such as human review enhance the quality of the data, ensuring it meets the desired standards. 

MongoDB~\cite{SAC2019,TSK+2019} is utilized in this scenario as a representative of NoSQL databases, which is consistent with the existing studies about text-to-NoSQL~\cite{LSQ+2025,QSL+2025}. Specifically, the relational databases in CoSQL~\cite{YZE+2019} were converted to NoSQL databases, and all SQL queries were transformed into NoSQL queries, with the help of TEND-based method~\cite{LSQ+2025}. The transformed NoSQL databases and the transformed NoSQL queries form the presented dataset~\DatasetName. Especially, the foreign key relationships between tables in the relational databases were transformed into the nested relationships in the NoSQL databases. $2$ problematic databases and $24$ associated dialogues were removed from \DatasetName. To verify the correctness of the transformed NoSQL query, the SQL query and the transformed NoSQL query are executed on SQLite database and MongoDB database, respectively. The transformed NoSQL query is considered correct, if the two results are identical (i.e., EVM=$1$).

However, only $60.13\%$ of the NoSQL queries transformed by the pipeline are correct. To further increase the percentage of the correct NoSQL queries, a RAG approach supported by multiple LLMs is proposed. 
Specifically, \DatasetName~is divided into two distinct segments according to the correctness of the transformed NoSQL queries. One subset, namely reference set, comprises the dialogues from which all incorrect question-query pairs have been removed. The other set contains all incorrect question-query pairs to refine. For example, given an incorrect NoSQL query, "db.Movie.find(\{\}, \{ title: 1, \_id: 0 \})", and its corresponding user question, "What is the name of the most recent movie", the pipeline will search the reference set for relevant question-query pairs by calculating the distance between the embedding vectors of the questions. The dialogues associated with these relevant pairs will then be retrieved and presented as similar dialogues. Following this, multiple LLMs are provided with the prompt that contains the user question, the incorrect NoSQL query, the NoSQL database information and the similar dialogues. The prompt is shown in Figure~\ref{prompt:dataset_regeneration} in Sec~\ref{sec:appendix_dataset} in the appendix. The objective of this approach is to empower the LLMs to re-generate the NoSQL query more accurately. After this process, the proportion of the correct NoSQL queries increases to $89.61\%$. 

\subsubsection{Dataset Statistics and Analysis}   
The statistics of {\DatasetName} dataset is discussed in this section. 
Table~\ref{tab:dataset_statistics} shows that {\DatasetName} dataset is  split into train and test set under cross domain setting. The average number of turns per dialogue is over $4$, revealing the semantic complexity of \DatasetName~dataset. Detailed statistics and analysis are provided in Section~\ref{sec:appendix_dataset} in the appendix.

\begin{table}[htbp]
    \centering
    \vspace{-5pt}
    \caption{Dataset Statistics}
    \vspace{-5pt}
    \scalebox{.95}{
    \begin{tabular}{@{}llll@{}}
    \toprule
                               & Train & Test  & Overall \\ \midrule
    Total \# Databases               & 139   & 19    & 158     \\
    Total \# Collections             & 335   & 37    & 372     \\
    Total \# Dialogue                & 2,115 & 276   & 2,391   \\
    Total \# turns             & 8,484 & 1,224 & 9,708   \\
    Avg. \# turns/Dialogues    & 4.01  & 4.43  & 4.06    \\
    Avg. \# Q.tokens/turns     & 11.91 & 11.79 & 11.90   \\
    Avg. \# NoSQL.tokens/turns & 31.38 & 32.98 & 31.58   \\
    Total \# Find Method       & 1,723 & 291   & 2,014   \\
    Total \# Aggregate Method  & 6,757 & 926   & 7,683   \\
    Total \# Other Method      & 4     & 7     & 11      \\ \bottomrule
    \end{tabular}
    }
    \vspace{-10pt}
    \label{tab:dataset_statistics}
\end{table}

%% file: sections/experiment.tex
\section{Experiment}   \label{sec:experiment}
\input{tables/comparison_table}

\subsection{Settings}
\subsubsection{Dataset}
\DatasetName~is used to evaluate the performance of the proposed method and baselines. To ensure its high quality, post-processing is applied: mismatched-execution queries are removed from the training set, while the test set undergoes manual verification (with both execution accuracy and semantic equivalence as metrics) via human intervention~\cite{WLX+2021, CKA2023}. Detailed information is provided in Section~\ref{subsec:appendix_quality_control} in the appendix.

\subsubsection{Base Models and Setup}
The state-of-the-art LLMs are employed to construct the baselines for evaluating the \DatasetName~dataset, including the open-source representatives such as DeepSeek-R1~\cite{GYZ+2025} and Qwen2.5-Coder-7B-Instruct~\cite{HYC+2024}, and the closed-source one, GPT-4o~\cite{HLG+2024}. 
Due to limited GPU resources, we performed three rounds of self-evolution exclusively on Qwen2.5-Coder-7B-Instruct for our proposed method. Detailed implementation steps and prompt templates are provided in Section~\ref{sec:appendix_exp} of the appendix.

\subsubsection{Baselines}
In our evaluation, we adopt several training-free paradigms, namely zero-shot learning~\cite{RLB2022,LHW+2023}, few-shot learning, and retrieval-augmented generation (RAG)~\cite{LPP+2020, GXG+2023, ZZY+2024}. The few-shot learning and RAG methods are under 10-shot setting.
To construct robust supervised fine-tuning (SFT) baselines, we implement two approaches: direct NoSQL prediction and the bootstrapped STaR-SFT method~\cite{ZWM+2022,PTS+2025}. Specifically, STaR-SFT is trained on the divide-and-conquer CoT reasoning data generated by DeepSeek-R1. Besides, we incorporate SMART~\cite{LSQ+2025}, a representative text-to-NoSQL method. We further report the performance of well-established agent-based methods, including ReAct~\cite{YZY+2023} and Plan-and-Solve~\cite{WXL+2023}. Detailed implementation steps and prompt templates are provided in Section~\ref{sec:appendix_exp} of the appendix.

\subsubsection{Metrics}
The metrics used in other text-to-NoSQL tasks~\cite{LSQ+2025, QSL+2025} are adopted in this experiment, including \emph{Exact Match (EM)} and \emph{Execution Accuracy (EX)}. \emph{EM} derives two metrics, \emph{Query Stages Match (QSM)} and \emph{Query Fields Coverage (QFC)}. There are also two subdivisions under \emph{EX}, \emph{Execution Fields Match (EFM)}, and \emph{Execution Value Match (EVM)}. In terms of evaluation focus, EM, QSM, and QFC assess the syntactic correctness of generated queries, whereas EX, EFM, and EVM verify their semantic correctness. Given the flexibility of NoSQL databases, where field names can be fully user-defined, we prioritize EVM over EX in our analysis to mitigate the interference caused by custom field naming conventions. Definitions of all metrics are provided in Section \ref{sec:appendix_exp_metrics} of the appendix.

\subsection{Main Experimental Results}

\subsubsection{Comparison}
The main results are shown in Table~\ref{tab:comparison}. Compared with other baseline models, {\MethodName} achieves superior performance in EVM accuracy, which serves as an effective indicator of semantic accuracy for NoSQL generation tasks. Notably, our method achieves a relative improvement of $6.7\%$ and $1.39\%$ in EVM accuracy over zero-shot large-scale models and RAG-based few-shot models, respectively. This result demonstrates that SLMs with 7 billion parameters are capable of addressing the challenging NoSQL generation task at a performance level comparable to that of much larger models, e.g., GPT-4o and DeepSeek-R1.

Regarding SFT-based methods, our approach, with and without the test-time scaling strategy (presented in Table~\ref{tab:comparison} and Table~\ref{tab:ablation_study_modules}, respectively), yields superior performance, which validates the effectiveness of the proposed stage-augmented CoT reasoning data.

However, agent-based methods achieve poor performance on the proposed task. One potential reason lies in their inherent lack of reasoning ability, which hinders them from constructing valid queries when faced with unseen NoSQL databases.

As an extension of our approach, we leverage a strong reasoning model, DeepSeek-R1, as the backbone to generate reasoning data in Round 1, a strategy that accelerates self-training and aligns with the findings of previous studies~\cite{GZL+2025}. In all subsequent self-training rounds, SLMs are deployed to reduce computational budget. This approach outperforms state-of-the-art large reasoning models, improving
EVM accuracy by up to $7.93\%$ under the zero-shot setting.

\subsubsection{Ablation Study}
An ablation study is conducted to explore the contribution of each module. The main results are presented in Table~\ref{tab:ablation_study_modules} and Table~\ref{tab:ablation_study_self_training}, with additional ablation results provided in Section~\ref{sec:appendix_exp_abation_study} of the appendix.

\noindent \textbf{Evaluation of Each Module.}
We validate the effectiveness of two core components, respectively: the Reward-based Sampling and Path Refinement (Reward-PR) component, and the MCTS-base Test-time Scaling (MCTS-TTS) component. All results are derived from three rounds of self-evolution.
Regarding the Reward-PR component ablated, we evaluate model performance using a rejection sampling strategy that only retains the reasoning paths for NoSQL queries with correct answers as the training set.
Regarding the MCTS-TTS component ablated, we set the number of output reasoning paths to 1. Table~\ref{tab:ablation_study_modules} demonstrates that the model performance degrades across all metrics, following the ablation of each individual component. This result verifies that every module contributes positively to the overall performance. 

\input{tables/ablation_study_modules}

\noindent \textbf{Evaluation of the Self-training Pipeline.}
Given the limited capacity of SLMs, we implement three rounds of MCTS-based in-depth reasoning to iteratively generate higher-quality data, and augment the training set with more challenging NoSQL generation samples. Key findings of the study are presented in Table~\ref{tab:ablation_study_self_training}. The base model without any fine-tuning is denoted as "SLM-base", with its round-1, round-2 and round-3 fine-tuned variants labeled "SLM-r1", "SLM-r2" and "SLM-r3", respectively. Specifically, we employ Qwen2.5-Coder-7B-Instruct as the representative SLM in the study. This ablation study demonstrates that the custom-designed self-training pipeline can effectively drive continuous performance improvement. Besides, the results in this table show that the growth rate of each metric decreases as the number of training rounds increases, indicating that the model performance is approaching convergence.

\input{tables/ablation_study_self_training}

\subsubsection{Case Study}
\begin{figure}[htbp]
    \centering
    \includegraphics[width=0.45\textwidth]{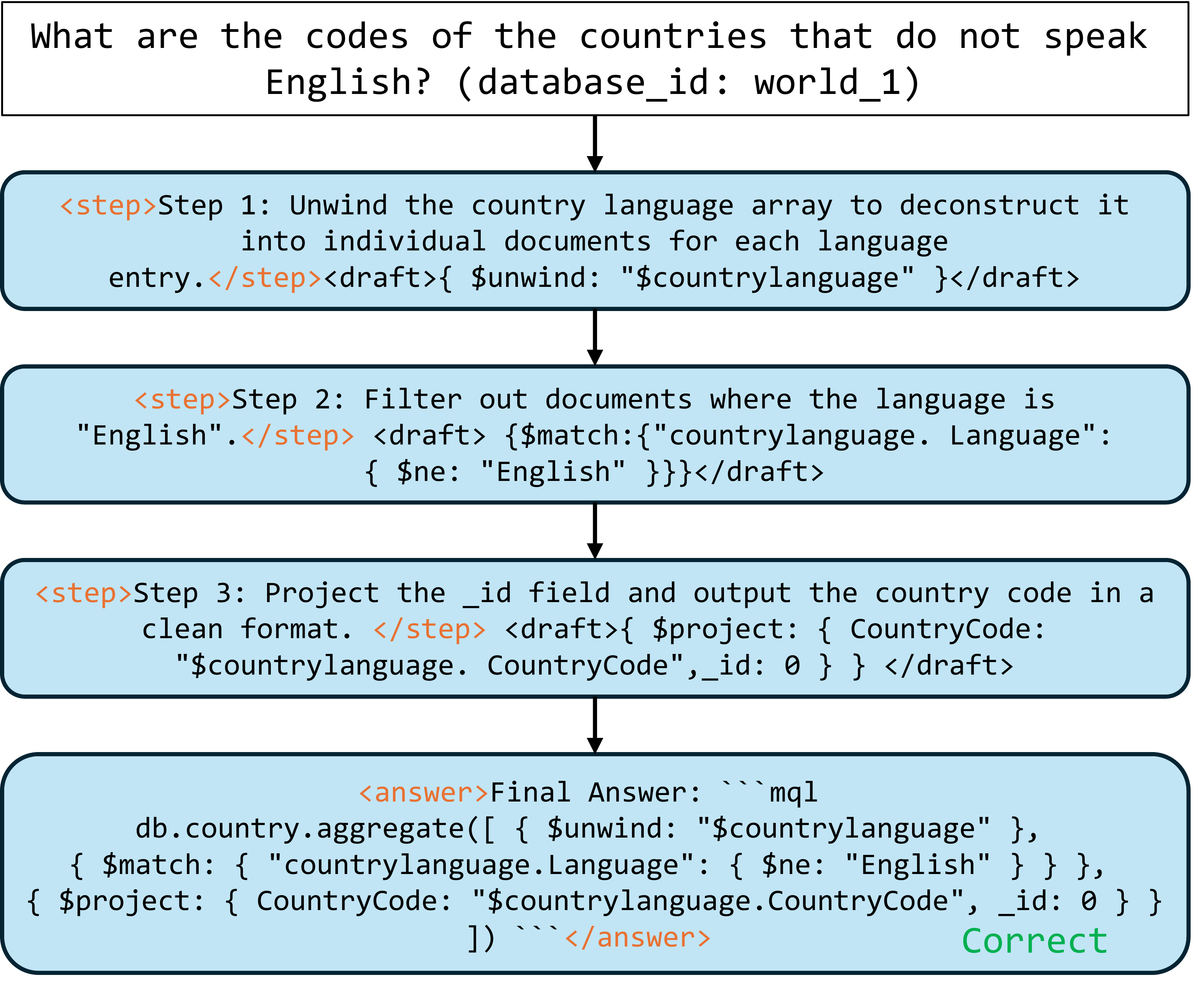}
    \vspace{-5pt}
    \caption{An example branch from the MCTS search tree (with nodes omitted for the page limit).}
    \vspace{-5pt}
    \label{fig:case_study_one_branch}
\end{figure}
Under our experimental setting, the MCTS search tree involves over 50 nodes and more than 10 distinct reasoning paths in total. Due to the page limit, we only showcase a single representative branch for illustrative purposes in this case study. As depicted in Figure~\ref{fig:case_study_one_branch}, the proposed method is capable of generating executable stages and the corresponding comments sequentially. Notably, it implicitly follows a plan-and-solve paradigm, where each reasoning step and the derived draft stage are grounded in the prior steps. For more comprehensive results, refer to Section~\ref{subsec:appendix_case_study} in the appendix. This supplementary result shows that our method can enumerate multiple candidate stages given identical prior steps, and then performs guided search based on the reward model.

\subsection{Summary}
{\MethodName} achieves superior EVM accuracy on the proposed task, outperforming zero-shot LLMs by up to $7.93\%$ and RAG-based
few-shot models by up to $2.62\%$. The results show that a 7B-parameter SLM is capable of matching the performance of larger-scale models, such as GPT-4o.  Ablation studies confirm the importance of both the Reward-PR and the MCTS-TTS components, and the self-training pipeline brings continuous improvement across rounds.

%% file: tables/comparison_table.tex
\begin{table*}[htbp]
\centering
\caption{Experimental results of different methods on six metrics. Bolded values stand for the best performance of all methods, with underlined values indicating the best performance of the baseline methods.}
\label{tab:comparison}
\scalebox{0.95}{
\begin{tabular}{lllcllcll}
\hline
 &
   &
   &
  \multicolumn{3}{c}{Query-based} &
  \multicolumn{3}{c}{Execution-based} \\ \hline
 &
  Method &
  Model Size &
  EM &
  \multicolumn{1}{c}{QSM} &
  \multicolumn{1}{c}{QFC} &
  EX &
  \multicolumn{1}{c}{EFM} &
  \multicolumn{1}{c}{EVM} \\ \hline
\multicolumn{1}{c}{\multirow{6}{*}{\rotatebox[origin=c]{90}{\begin{tabular}[c]{@{}c@{}}Training-free\\ Paradigms\end{tabular}}}} &
  Zero-Shot (GPT-4o) &
  UNK &
  0.1413 &
  \multicolumn{1}{c}{0.4649} &
  \multicolumn{1}{c}{0.6242} &
  0.4248 &
  \multicolumn{1}{c}{0.5433} &
  \multicolumn{1}{c}{0.5654} \\
\multicolumn{1}{c}{} &
  Zero-Shot (DeepSeek-R1) &
  671B &
  0.1144 &
  \multicolumn{1}{c}{0.4575} &
  \multicolumn{1}{c}{0.6495} &
  0.4779 &
  \multicolumn{1}{c}{0.5817} &
  \multicolumn{1}{c}{0.6168} \\
\multicolumn{1}{c}{} &
  Few-Shot (GPT-4o) &
  UNK &
  {\ul \textbf{0.1716}} &
  \multicolumn{1}{c}{0.5155} &
  \multicolumn{1}{c}{0.6667} &
  0.4967 &
  \multicolumn{1}{c}{0.6078} &
  \multicolumn{1}{c}{0.6577} \\
\multicolumn{1}{c}{} &
  Few-Shot (DeepSeek-R1) &
  671B &
  0.1601 &
  \multicolumn{1}{c}{0.5033} &
  \multicolumn{1}{c}{0.6634} &
  0.5106 &
  \multicolumn{1}{c}{0.6176} &
  \multicolumn{1}{c}{0.6593} \\
\multicolumn{1}{c}{} &
  RAG (GPT-4o) &
  UNK &
  0.1699 &
  \multicolumn{1}{c}{{\ul \textbf{0.5261}}} &
  \multicolumn{1}{c}{{\ul \textbf{0.6683}}} &
  0.5033 &
  \multicolumn{1}{c}{0.6152} &
  \multicolumn{1}{c}{0.6503} \\
\multicolumn{1}{c}{} &
  RAG (DeepSeek-R1) &
  671B &
  0.1422 &
  \multicolumn{1}{c}{0.5155} &
  \multicolumn{1}{c}{0.6618} &
  {\ul 0.5270} &
  \multicolumn{1}{c}{{\ul \textbf{0.6299}}} &
  \multicolumn{1}{c}{{\ul 0.6699}} \\ \hline
\multirow{2}{*}{\rotatebox[origin=c]{90}{SFT}} &
  SFT-direct (Qwen2.5-Coder) &
  7B &
  \multicolumn{1}{l}{0.1511} &
  0.5008 &
  0.5735 &
  \multicolumn{1}{l}{0.4730} &
  0.5572 &
  0.5899 \\
 &
  STaR-SFT~\cite{ZWM+2022} (Qwen2.5-Coder, DeepSeek-R1) &
  7B &
  \multicolumn{1}{l}{0.1021} &
  0.4461 &
  0.6373 &
  \multicolumn{1}{l}{0.4649} &
  0.5531 &
  0.6454 \\ \hline
\multirow{3}{*}{\rotatebox[origin=c]{90}{Agent}} &
  ReAct~\cite{YZY+2023} (GPT-4o) &
  UNK &
  \multicolumn{1}{l}{0.0891} &
  0.3954 &
  0.5605 &
  \multicolumn{1}{l}{0.3587} &
  0.4706 &
  0.4559 \\
 &
  Plan-and-Solve~\cite{WXL+2023} (GPT-4o) &
  UNK &
  \multicolumn{1}{l}{0.0760} &
  0.3734 &
  0.6038 &
  \multicolumn{1}{l}{0.4175} &
  0.4918 &
  0.5417 \\
 &
  SMART~\cite{LSQ+2025} (GPT-4o) &
  UNK &
  \multicolumn{1}{l}{0.1111} &
  0.4902 &
  0.5931 &
  \multicolumn{1}{l}{0.4583} &
  0.5400 &
  0.5654 \\ \hline
\multirow{2}{*}{\rotatebox[origin=c]{90}{Ours}} &
  Stage-MCTS (Qwen2.5-Coder) &
  7B &
  \multicolumn{1}{l}{0.1462} &
  0.5221 &
  0.6552 &
  \multicolumn{1}{l}{\textbf{0.5392}} &
  0.6266 &
  0.6838 \\
 &
  Stage-MCTS (Qwen2.5-Coder, DeepSeek-R1) &
  7B &
  \multicolumn{1}{l}{0.1332} &
  0.5049 &
  0.6528 &
  \multicolumn{1}{l}{0.5335} &
  0.6144 &
  \textbf{0.6961} \\ \hline
\end{tabular}
}
\end{table*}

%% file: tables/ablation_study_modules.tex
\begin{table}[htbp]
\centering
\caption{Ablation study on each module after three rounds of self-training.}
\label{tab:ablation_study_modules}
\scalebox{.9}{
\begin{tabular}{lllllll}
\hline
              & \multicolumn{3}{c}{Query-based} & \multicolumn{3}{c}{Execution-based} \\ \hline
Method        & EM        & QSM      & QFC      & EX         & EFM        & EVM       \\ \hline
StageMCTS     & 0.1462    & 0.5221   & 0.6552   & 0.5392     & 0.6266     & 0.6838    \\
w/o Reward-PR & 0.0629    & 0.3930   & 0.5792   & 0.4526     & 0.5368     & 0.6176    \\
w/o MCTS-TTS  & 0.1413    & 0.5123   & 0.6479   & 0.5180     & 0.6225     & 0.6601    \\ \hline
\end{tabular}
}
\end{table}

%% file: tables/ablation_study_self_training.tex
\begin{table}[htbp]
\centering
\caption{Performance of the self-training SLM in each round, which shows continuous improvement.
}
\label{tab:ablation_study_self_training}
\scalebox{0.9}{
\begin{tabular}{lllllll}
\hline
         & \multicolumn{3}{c}{Query-based} & \multicolumn{3}{c}{Execution-based}        \\ \hline
Round\# & \multicolumn{1}{c}{EM} & \multicolumn{1}{c}{QSM} & \multicolumn{1}{c}{QFC} & \multicolumn{1}{c}{EX} & \multicolumn{1}{c}{EFM} & \multicolumn{1}{c}{EVM} \\ \hline
SLM-base & 0.0882    & 0.4232   & 0.5580   & 0.3905          & 0.4771          & 0.5008 \\ \hline
SLM-r1   & 0.1250    & 0.4935   & 0.6520   & 0.5270          & 0.6201          & 0.6675 \\
SLM-r2   & 0.1397    & 0.5106   & 0.6503   & \textbf{0.5433} & \textbf{0.6266} & 0.6806 \\
SLM-r3  & \textbf{0.1462}        & \textbf{0.5221}         & \textbf{0.6552}         & 0.5392                 & \textbf{0.6266}         & \textbf{0.6838}         \\ \hline
\end{tabular}
}
\end{table}

%% file: sections/related_work_short.tex
\section{Related Work}  \label{sec:related_work}
This study is closely related to the field of user-friendly NoSQL database, text-to-SQL and LLM reasoning.

\smallskip

\noindent \textbf{User Friendly NoSQL Database.}

Not only SQL (NoSQL) databases are advanced data storage and management systems that store data in non-tabular formats without a pre-defined structure. In databases and data mining, existing research on NoSQL databases mainly focuses on several key areas. For example, study~\cite{C2011} discussed the availability and scalability of NoSQL database systems. Another paper~\cite{OGG+2011} examined security and privacy issues in NoSQL databases. Additionally, research~\cite{MBZ+2019} aims to improve the energy efficiency of NoSQL databases through query optimizations. Moreover, effort~\cite{AED+2011}  was devoted to achieving scalability, elasticity, and autonomy in database management systems within cloud computing setups.

Despite extensive NoSQL research across domains, existing work overlooks accessibility for non-expert users lacking NoSQL query knowledge. To address this, a healthcare-focused text-to-ESQ (Elasticsearch Query) task~\cite{ZZY+2023} was proposed, enabling the precise and efficient exploration of complex healthcare data in NoSQL environments. Besides, the study~\cite{ZHY+2024} presented an explainable chain-of-thought method to leverage LLMs for the text-to-ESQ task. Recently, the work~\cite{LSQ+2025} proposed a text-to-NoSQL task for natural language interaction with NoSQL databases, constructing the TEND dataset (for MongoDB queries) to support cross-domain partitioning. MultiTEND~\cite{QSL+2025} was subsequently introduced as the largest multilingual benchmark for multilingual text-to-NoSQL tasks.

However, the previous work~\cite{ZZY+2023,ZHY+2024,LSQ+2025,QSL+2025} focused on converting a single question into its corresponding NoSQL query, while the task of conversational text-to-NoSQL is proposed for mapping the context-dependent questions to the associated NoSQL query.

\smallskip

\noindent \textbf{Text-to-SQL.}
Text-to-SQL is significant in reducing the barrier to accessing the relational database~\cite{KK2023, HYZ+2024, LSL+2024}. Early efforts~\cite{PAE+2004, SKD2005, LJ2014} rely primarily on the rule-based methods. They are implemented based on well-designed rules and templates, especially suitable for simple scenarios. Subsequently, deep neural networks~\cite{ZXS2017,XLS2017,YLZ+2018,CSK+2021} and pre-trained language models (PLMs)~\cite{DCL+2019, RSR+2020} have been developed for text-to-SQL tasks, achieving promising results. Specifically, TaBERT~\cite{YNY+2020} was trained on a large corpus of the tabular data and the English context, in order to jointly understand the textual and tabular data. RESDSQL~\cite{LZL+2023} was designed as a ranking-enhanced encoding and skeleton-aware decoding framework, which improved the accuracy and robustness by decoupling the schema linking and the skeleton parsing. Graphix-T5~\cite{LHC+2023} enhanced the pre-trained T5 model~\cite{RSR+2020} by incorporating specially-designed graph-aware layers, which were intended to increase the reasoning capacity of model. For the purpose of evaluating how effectively text-to-SQL models operate in practical scenarios, multiple extensive benchmark datasets have been developed and released, including WikiSQL~\cite{ZXS2017}, Spider~\cite{YZY+2018}, SParC~\cite{YZY+2019}, CoSQL~\cite{YZE+2019}, BIRD~\cite{LHQ+2023} and Spider 2.0~\cite{LCY+2025}.

Recently, there has been a growing interest in large language model (LLM) approaches. LLM-based method have implemented text-to-SQL through in-context learning~\cite{PR2023, GWL+2024, TPC+2024, RFH+2024, SXL+2025, DRX+2025,TSC2025} and supervised fine-tuning (SFT)~\cite{LZL+2024,PR2024,GGL+2025,QCF+2025} paradigms, achieving the state-of-the-art accuracy with well-designed frameworks and stronger comprehension capabilities.

However, the proposed conversational text-to-NoSQL task is different from text-to-SQL task, especially for SParC and CoSQL. For example, conversational text-to-NoSQL aims to retrieve the data from the collections, while text-to-SQL focuses on tables. When handling complex questions, text-to-SQL has to generate queries that can conduct joint searches across different relational tables. In contrast, conversational text-to-NoSQL solves this problem by querying a single document with nested fields. Besides, the schema of NoSQL databases is more flexible than that of SQL databases, making the task of conversational text-to-NoSQL more complex. 

\smallskip

\noindent \textbf{Large Language Model Reasoning.}
With the help of the Chain-of-Thought (CoT) technique~\cite{WWS+2022,KGR+2022}, LLMs have achieved remarkable performance improvements across a wide range of tasks~\cite{GZL+2025,HGM+2023,SSR+2025}. As a representative reasoning paradigm, CoT mimics human-like cognitive processes by decomposing complex questions into sequential reasoning steps to derive coherent solutions. Building on this foundation, the Tree-of-Thoughts (ToT) framework~\cite{YYZ+2023} further generalizes the CoT methodology. It constructs a tree-structured reasoning space that encompasses multiple alternative reasoning paths and explores semantically coherent thought units to enable deliberate and thoughtful decision-making.

Existing research on LLM reasoning primarily focuses on synthesizing large-scale, diverse datasets~\cite{YJS+2024,GSY+2025,WHL+2024,SHZ+2024} and designing more effective algorithms to enhance the reasoning capabilities of LLMs~\cite{ZWM+2022,ZZH+2024,YML+2025,TDG+2025}. However, most of the studies center on the math reasoning and science tasks. To address this gap, this work represents the first attempt to advance the reasoning capabilities of LLMs in the database domain, leveraging an improved MCTS-based data augmentation approach and a tailored self-training pipeline.

%% file: sections/conclusion.tex
\section{Conclusion}    
\label{sec:conclusion}
In this work, we identified the limitations of single-turn text-to-NoSQL systems and proposed the novel conversational text-to-NoSQL task to enable multi-turn and context-aware query generation. We introduced \MethodName, a framework that formulates NoSQL query generation as a search problem and equips SLMs with NoSQL-specific reasoning through MCTS-based data augmentation, progressive fine-tuning, and iterative self-training. To support this research, we built \DatasetName, a large-scale cross-domain dataset constructed via LLM-human collaboration. Extensive experiments validated the effectiveness of our approach, which outperforms advanced models like DeepSeek-R1 and GPT-4o in EVM accuracy.

Future work will focus on integrating the conversational query system into  broader data exploration workflows, including automated visualization generation and deep research agents.

%% file: sections/appendix.tex
\clearpage

\appendix

\input{appendices/preliminary_experiment}
\input{appendices/datasets}

\input{appendices/experiments}

\input{appendices/prompts}

%% file: appendices/preliminary_experiment.tex
\section{Preliminary Experiments and Method Details}   \label{sec:appendix_methods}
\subsection{Reasoning Capabilities of SLMs for NoSQL Databases}
In this section, Qwen2.5-Coder-7B-Instruct is adopted as the backbone model.
Table~\ref{tab:preliminary_exp_NoSQL_reasoning} demonstrates that the performance of SLMs generating NoSQL queries with the assistance of intermediate reasoning steps is inferior to that of SLMs generating such queries directly. This phenomenon can be attributed to the inherent lack of NoSQL-specific reasoning capabilities in SLMs. Zero-Shot-Direct refers to the setting where the SLM directly generates NoSQL queries, whereas Zero-Shot-CoT denotes the setting where the model generates NoSQL queries only after producing intermediate reasoning steps. Details of the prompt design can be found in Figure~\ref{prompt:star_sft}.

\begin{table}[htbp]
\centering
\caption{Preliminary Experiment on NoSQL-Oriented Reasoning Capabilities of SLMs.}
\label{tab:preliminary_exp_NoSQL_reasoning}
\scalebox{0.9}{
\begin{tabular}{lllllll}
\hline
                 & \multicolumn{3}{c}{Query-based} & \multicolumn{3}{c}{Execution-based} \\ \hline
Method           & EM        & QSM      & QFC      & EX         & EFM        & EVM       \\ \hline
Zero-Shot-Direct & 0.0882    & 0.4232   & 0.5580   & 0.3905     & 0.4771     & 0.5008    \\
Zero-Shot-CoT    & 0.0376    & 0.2868   & 0.5016   & 0.3194     & 0.3775     & 0.3913    \\ \hline
\end{tabular}
}
\end{table}

\subsection{Additional Method Details}

\subsubsection{Path Refinement}
The prompt for path refinement via SLM is presented in Figure~\ref{prompt:path_refinement}. Specifically, we only retain the content enclosed within the <step> and <draft> tags. The content of the <reasoning> tag is discarded, as it serves solely to enhance the performance of the SLM in refining these paths and is not required for subsequent processing. For the sake of training efficiency, we set the top-$k$ selection to top-1.

\subsubsection{Self-training Pipeline} \label{subsubsec:appendix_self-training}
Based on our extensive experiments, the SLM is fine-tuned from the initial base model in each round, rather than training incrementally on the model from the previous round to alleviate the risk of overfitting. This training design aligns with the findings of recent study~\cite{GZL+2025}.
Detailed prompt templates corresponding to the three-phase supervised fine-tuning process are presented in Figure~\ref{prompt:generate_next_step} and Figure~\ref{prompt:rollout}, respectively.

%% file: appendices/datasets.tex
\begin{table*}[htbp]
\centering
\caption{Comparison of \DatasetName~with other existing Text-to-Query datasets.}
\begin{tabular}{@{}lcccc@{}}
\toprule
Dataset   & Database Type & Target Language & Construction & Multi-turn \\ \midrule
VAERSESQ~\cite{ZZY+2023} & ElasticSearch & NoSQL & Synthesis    & No      \\
TEND~\cite{LSQ+2025}    & MongoDB   & NoSQL    & Manual+LLM   & No      \\
MultiTEND~\cite{QSL+2025} & MongoDB   & NoSQL    & Manual+LLM   & No      \\
SParC~\cite{YZY+2019}     & SQLite   & SQL     & Manual       & Yes     \\
CoSQL~\cite{YZE+2019}     & SQLite   & SQL     & Manual       & Yes     \\ \midrule
CoNoSQL~(Ours)    & MongoDB   & NoSQL    & Manual+LLM   & Yes     \\ \bottomrule
\end{tabular}

\label{tab:dataset_comparison}

\end{table*}

\section{Additional Dataset Details}   \label{sec:appendix_dataset}

\subsection{CoSQL Datasaet}
CoSQL~\cite{YZE+2019}, a popular and complex conversational text-to-SQL corpus under cross-domain setting, was used in this experiment. The cross-domain setting refers to the situation where the specific domains (e.g., restaurants and flights) among the training set, the development set, and the test set are different. The training set and the development set of CoSQL were utilized to construct the dataset. Notably, the development set of CoSQL is regarded as the test set of CoNoSQL. These sets collectively contain a total of $166$ databases and $2,459$ dialogues. Each dialogue comprises multiple user questions and the corresponding SQL queries. The number of dialogue turns, i.e., question-SQL pairs, is $10,802$.

\subsection{Dataset Construction}
Table~\ref{tab:dataset_comparison} clearly demonstrates the key differences between the presented dataset and those utilized in prior relevant studies. Especially, unlike the existing conversational text-to-SQL dataset (e.g., SParC and CoSQL), the proposed dataset concentrates on a schema that is more flexible than the pre-defined structure of SQL databases. Besides, for complex questions, the SQL queries in the previous datasets typically involve cross-table joins within relational databases. In contrast, the NoSQL queries simplify this by querying a single document with nested fields in NoSQL databases.  
Figure~\ref{prompt:dataset_regeneration} presents the prompt employed for the refinement and regeneration of the proposed dataset, \DatasetName.

As for the settings of the approach, GPT-4o, DeepSeek-R1~\cite{GYZ+2025} and Llama3-70B~\cite{DJP+2024} were employed as the representatives of LLMs. The number of the similar dialogues was set to 3. Each dialogue should be different. For the embeddings of the questions, this experiment adpoted the pretrained model, all-MiniLM-L6-v2, from the HuggingFace Sentence Transformers library~\cite{WDS+2019}.

\subsection{Quality Control}    \label{subsec:appendix_quality_control}
To further ensure the high quality of the \DatasetName~dataset, post-processes are employed to improve the quality. For the training set, the queries whose execution results do not match those of the corresponding SQL were removed. For the test set, human intervention~\cite{WLX+2021, CKA2023} is used. Specifically, all the queries were checked manually, where the measurements are not only the execution accuracy, but also the semantic equivalence. For instance, consider a user question asking, "Can I see the list of customers?" The SQL query "Select * From Customers" and the transformed NoSQL query "db.Customers.find()" are re-labeled as semantically equivalent, even though their execution results differ, which arises from the distinct schema inherent in the SQL and NoSQL databases. Thus, it is necessary to re-label the transformed NoSQL queries through human evaluation. 
The evaluation criteria are as follows.
\begin{itemize}
    \item Coherent (True or False): Are the NoSQL queries relevant and coherent with the dialogue context?
    \item Faithfulness (True or False): Are the output of the NoSQL queries consistent with the user questions?
\end{itemize}

After being manually evaluated, the remaining incorrect NoSQL queries will be corrected by the experts (i.e., the authors of the paper). The summary of manual correction is as follows.
\begin{itemize}
    \item The aggregate method like "db.teacher.aggregate([ \{ \$count: "count" \} ])" is preferred to used to count the number of selected documents, since the execution result of the semantically equivalent NoSQL "db.teacher.countDocuments({})" will be an empty set instead of “0” if the number of selected documents is 0.
    \item The condition "preserveNullAndEmptyArrays: true" is used to avoid the error that the execution results of some cases are empty arrays.
    \item Before using the aggregate operators such as "sum" and "count", the "\$toDouble" operator is used to avoid the error of counting.
    \item The semantic equivalence is considered in the manual review, even though the execution result of the SQL query is different from that of the transformed NoSQL query.
\end{itemize}

\subsection{Dataset Statistics and Analysis}    
The statistics of {\DatasetName} dataset is discussed in this section. Table~\ref{tab:database_statistics} summarizes the statistics of the NoSQL databases. The average number of the fields per database is $15.09$ and that of the documents per database is over $800$, indicating the complexity of the databases.
Table~\ref{tab:method_statistics} concludes the statistics of the methods used in the NoSQL queries of both the training set and the test set. It shows that the operations/stages used in the NoSQL queries of \DatasetName~dataset have wide coverage and diversity.

\begin{table}[]
    \centering
    \scalebox{.85}{
    \begin{tabular}{|cccc|}
    \hline
    \multicolumn{4}{|c|}{The number of Fields and Documents}                                    \\ \hline
    \multicolumn{1}{|c|}{\# Fields} & \multicolumn{1}{c|}{Avg. \# Fields/DBs} & \multicolumn{1}{c|}{Max. \# Fields} & Min. \# Fields \\ \hline
    \multicolumn{1}{|c|}{2,475}   & \multicolumn{1}{c|}{15.09}  & \multicolumn{1}{c|}{53}    & 2 \\ \hline
    \multicolumn{1}{|c|}{\# Docs}   & \multicolumn{1}{c|}{Avg. \# Docs/DBs}   & \multicolumn{1}{c|}{Max. \# Docs}   & Min. \# Docs   \\ \hline
    \multicolumn{1}{|c|}{132,469} & \multicolumn{1}{c|}{807.74} & \multicolumn{1}{c|}{44,139} & 0 \\ \hline
    \end{tabular}
    }
    \caption{Database Statistics}
    \label{tab:database_statistics}
\end{table}

\begin{table}[]
    \centering
    \scalebox{.85}{
    \begin{tabular}{|ccccc|}
    \hline
    \multicolumn{5}{|c|}{Operations in Find Method}                                                                                                 \\ \hline
    \multicolumn{1}{|c|}{\# Project} & \multicolumn{1}{c|}{\# Match}   & \multicolumn{1}{c|}{\# Sort}  & \multicolumn{1}{c|}{\# Limit}  & \# Others \\ \hline
    \multicolumn{1}{|c|}{2,003}      & \multicolumn{1}{c|}{1,383}      & \multicolumn{1}{c|}{370}      & \multicolumn{1}{c|}{319}       & 108       \\ \hline
    \multicolumn{5}{|c|}{Stages in Aggregate Method}                                                                                                \\ \hline
    \multicolumn{1}{|c|}{\# Unwind}  & \multicolumn{1}{c|}{\# Project} & \multicolumn{1}{c|}{\# Match} & \multicolumn{1}{c|}{\# Group}  & \# Sort   \\ \hline
    \multicolumn{1}{|c|}{9,277}      & \multicolumn{1}{c|}{7,374}      & \multicolumn{1}{c|}{5,649}    & \multicolumn{1}{c|}{4,160}     & 1,548     \\ \hline
    \multicolumn{1}{|c|}{\# Lookup}  & \multicolumn{1}{c|}{\# Limit}   & \multicolumn{1}{c|}{\# Count} & \multicolumn{1}{c|}{\# Others} &           \\ \hline
    \multicolumn{1}{|c|}{1,491}      & \multicolumn{1}{c|}{1,394}      & \multicolumn{1}{c|}{780}      & \multicolumn{1}{c|}{3,230}     &           \\ \hline
    \end{tabular}
    }
    \caption{Method Statistics}
    \label{tab:method_statistics}
\end{table}

%% file: appendices/experiments.tex
\section{Additional Experiments and Details}   \label{sec:appendix_exp}

\subsection{Implementation Details}
\subsubsection{Baselines}
The prompt employed for training-free paradigms is detailed in Figure~\ref{prompt:in_context_learning}. 
For the supervised fine-tuning baselines, direct NoSQL prediction adheres to the Alpaca format provided in LLaMA-Factory~\cite{ZZZ+2024}. The prompt used for the bootstrapped STaR-SFT method is detailed in Figure~\ref{prompt:star_sft}.
For the single-turn text-to-NoSQL model, SMART, we concatenated the user question and dialogue history into a unified text sequence. This sequence was then used to construct prompts, which were subsequently fed into the language models for processing. 
Agent-based methods were implemented using the LangChain \footnote{https://www.langchain.com/} platform, with their corresponding prompts detailed in Figure~\ref{prompt:react} and Figure~\ref{prompt:plan_and_solve}, respectively.

\subsubsection{Hyperparameter}
For the training-free paradigms, a temperature of 0.0 was adopted, as with most studies~\cite{LCY+2025,YJW+2024,CLZ+2024}. The input text will be truncated from the beginning if it exceeds the max token limit required by the models. The few-shot learning and RAG methods are under 10-shot setting. In the few-shot learning setting, similar examples are selected at random.
For the supervised fine-tuning (SFT) method, the number of training epochs is $3$ with a batch size of $32$. We use AdamW optimizer with a cosine learning rate scheduler, setting the initial learning rate to "1.0e-5". Detailed configuration for {\MethodName} is presented in Table~\ref{tab:method_hyperparameter}.

\begin{table}[htbp]
\caption{Hyperparameters Used in \MethodName}
\begin{tabular}{@{}ll@{}}
\toprule
Hyperparameter & Value \\ \midrule
$c$              & 1.414 \\
$m$              & 3     \\
$max\_depth$     & 8     \\
$N_{rollout}$    & 10    \\
top-$k$          & 1     \\ \bottomrule
\end{tabular}
\label{tab:method_hyperparameter}
\end{table}

\begin{figure*}[htbp]
    \centering
    \includegraphics[width=0.99\textwidth]{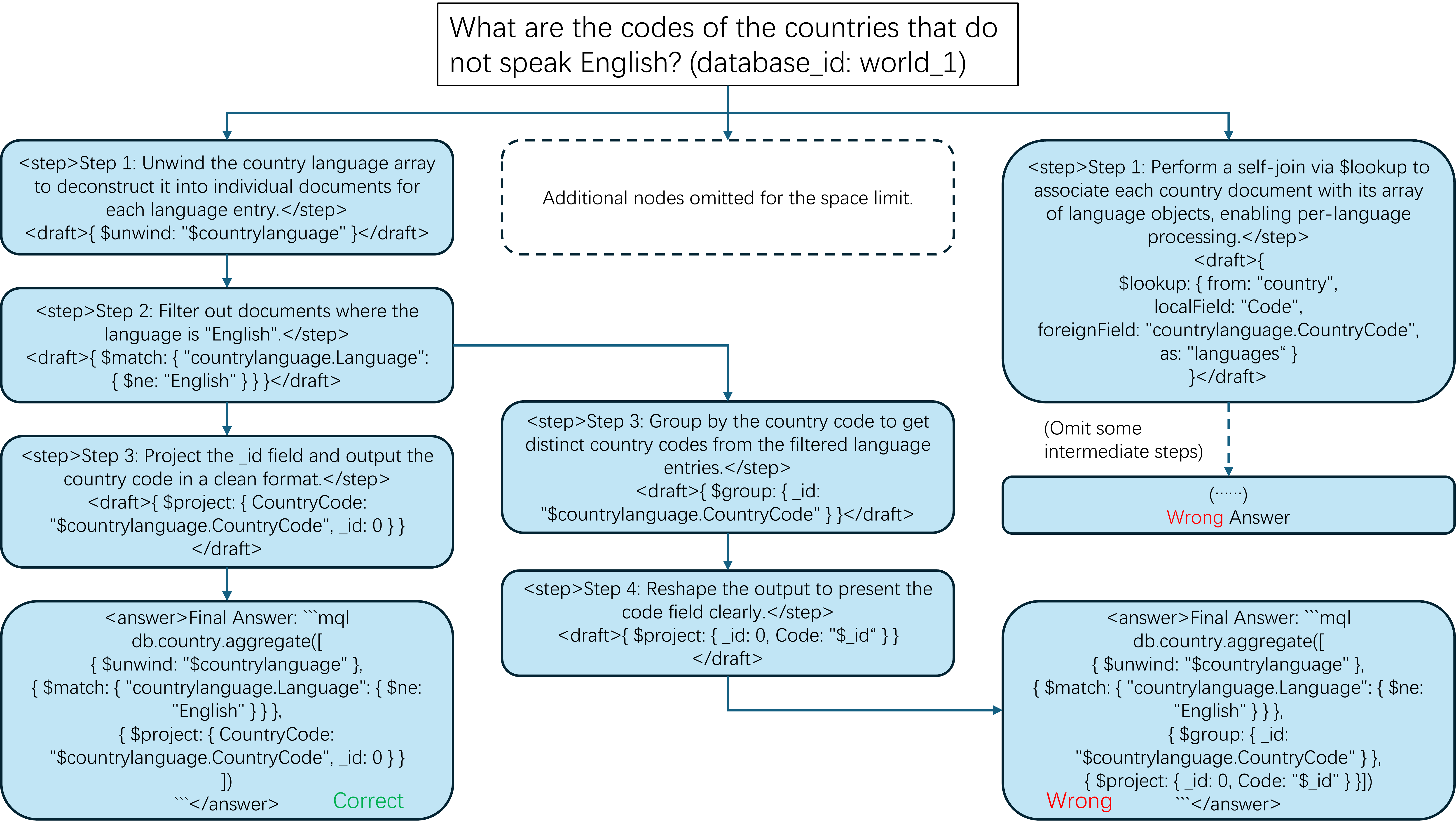}
    \caption{An example of the MCTS-based NoSQL query generation. This supplementary figure shows that our method can enumerate multiple candidate stages given identical prior steps, and then performs guided search based on the reward model.}
    \label{fig:case_study_full}
\end{figure*}

\subsection{Metrics}    \label{sec:appendix_exp_metrics}
The detailed definitions of the metrics are as follows.

\subsubsection{Exact Match (EM)}
EM is designed to evaluate whether the predicted query is an exact match to the ground-truth query. The calculation formula is
$$EM=\frac{N_{em}}{N},$$
where $N_{em}$ denotes the number of queries fully matching the ground-truth, and $N$ represents the total number of the queries within the test set.

\subsubsection{Query Stages Match (QSM)}
QSM is to assess whether the key stages within the predicted NoSQL query align with those in the ground-truth NoSQL query, in terms of both the order and the stage keywords. These stages comprise match, group, lookup, etc. The calculation formula is
$$QSM=\frac{N_{qsm}}{N},$$
where $N_{qsm}$ is defined as the number of queries, in which the key stages of the predicted query match exactly with those in the ground-truth query.

\subsubsection{Query Fields Coverage (QFC)}
QFC is utilized to evaluate whether the fields in the predicted query comprehensively cover all the fields present in the ground-truth query. This includes both database fields and query-defined fields. The calculation formula is
$$QFC=\frac{N_{qfc}}{N},$$
where $N_{qfc}$ represents the number of queries that completely cover all the fields present in the corresponding ground-truth queries.

\subsubsection{Execution Accuracy (EX)}
EX is designed to measure the accuracy of the execution results for the predicted queries on the database. The calculation formula is
$$EX=\frac{N_{ex}}{N},$$
where $N_{ex}$ denotes the number of queries whose execution results are exactly equal to those of their corresponding ground-truth queries.

\subsubsection{Execution Fields Match (EFM)}
EFM is to assess whether the field names within the execution results of the predicted query align precisely with those in the expected results, which are the execution outcomes of the ground-truth query. The calculation formula is
$$EFM=\frac{N_{efm}}{N},$$
where $N_{efm}$ is the number of queries with completely matching field names in the execution results.

\subsubsection{Execution Value Match (EVM)}
EVM is utilized to evaluate whether the values in the execution results of the predicted query match those of the ground-truth query. The calculation formula is
$$EVM=\frac{N_{evm}}{N},$$
where $N_{evm}$ is the number of queries with fully matching values in the execution results.

\input{tables/ablation_study_modules_all_rounds}

\input{tables/case_study215}

\subsection{Additional Ablation Study} \label{sec:appendix_exp_abation_study}

\noindent \textbf{Evaluation of the Self-training Pipeline.}
Table~\ref{tab:ablation_study_modules_all_rounds} presents a detailed ablation study of each module across all self-training rounds. The table shows that model performance degrades across all metrics when any individual component is ablated, verifying that each module contributes positively to the overall performance. Notably, the performance of all models improves with an increasing number of rounds, demonstrating the effectiveness of the proposed self-training pipeline.

\noindent \textbf{Evaluation of the Test-time Scaling Strategy.}
This ablation study is conducted to evaluate the effectiveness of the reward model within the test-time scaling strategy. The main results are presented in Table~\ref{tab:ablation_study_reward_model}. In {\MethodName}, the self-supervised reward model (SRM) is integrated with MCTS. For comparison, Self-Consistency and Best-of-N (BoN)~\cite{LKB+2024} are also adopted as scaling methods under the same search budget as the proposed approach, with Self-Consistency using the same EVM metric as ours.
The process reward model (PRM) employed in this ablation study is identical to that in \cite{ZZH+2024}. Specifically, reasoning paths are partitioned into two categories, correct and incorrect, with ground-truth labels of 1.0 and -1.0 assigned, respectively. PRM training is formulated as a classification task, and the model is constructed by appending a linear layer to the policy SLM to directly map output probabilities to a scalar reward value. Notably, these results validate the effectiveness of the proposed method: SRM is simple yet highly effective. Both test-time scaling methods augmented with PRM yield worse performance than the baseline without any test-time scaling, indicating that PRM fails to accurately judge the correctness of generated NoSQL queries, due to the inherent complexity of NoSQL generation.
Regarding Self-Consistency, it primarily improves query-based performance, but its execution-based metrics are lower than those of MCTS with SRM.

\input{tables/ablation_study_reward_model}

\subsection{Additional Case Study}  \label{subsec:appendix_case_study}
Figure~\ref{fig:case_study_full} shows that {\MethodName} can enumerate multiple candidate stages given identical prior steps, and then performs guided search based on the reward model.

Table~\ref{tab:case_study_215} and Table~\ref{tab:case_study_182} present concrete examples that illustrate the NoSQL queries and corresponding execution results generated by the single-path baselines and our proposed tree-based search method, \MethodName. Specifically, Table~\ref{tab:case_study_215} illustrates that when generating simple queries, {\MethodName} is able to simplify the aggregation pipeline by using only the single \$count operator. In contrast, the zero-shot approach augmented by GPT-4o commits the same error as STaR-SFT. It returns $15$ professionals rather than the correct value of $8$. Both methods incorrectly apply \$size to the Treatments array, which fails to accurately reflect the actual number of treatments. Table~\ref{tab:case_study_182} shows that for complex query generation, {\MethodName} adheres to MongoDB best practices and employs appropriate operators. However, the GPT-4o-augmented zero-shot approach fails due to a data conversion error, indicating its inability to handle potential data quality issues. As for STaR-SFT, although it yields the correct answer (i.e., 'amc'), it additionally returns incorrect answers (i.e., 'renault' and 'ford') and relies on an unnecessarily complex aggregation pipeline with redundant grouping and projection stages.

\input{tables/case_study182}

%% file: tables/ablation_study_modules_all_rounds.tex
\begin{table}[htbp]
\centering
\caption{Ablation study on each module across all rounds of self-training.}
\label{tab:ablation_study_modules_all_rounds}
\scalebox{.9}{
\begin{tabular}{lllllll}
\hline
              & 
\multicolumn{3}{c}{Query-based} & \multicolumn{3}{c}{Execution-based} \\ \hline
Method        & EM        & QSM      & QFC      & EX         & EFM        & EVM       \\ 
\hline
\multicolumn{7}{c}{Round 1}                                                           \\ \hline
StageMCTS     & 0.1250    & 0.4935   & 0.6520   & 0.5270     & 0.6201     & 0.6675    \\
w/o Reward-PR & 0.0596    & 0.3734   & 0.5572   & 0.4183     & 0.4943     & 0.5605    \\
w/o MCTS-TTS  & 0.1176    & 0.4796   & 0.6438   & 0.4894     & 0.6005     & 0.6315    \\ 
\hline
\multicolumn{7}{c}{Round 2}                                                           \\ \hline
StageMCTS     & 0.1397    & 0.5106   & 0.6503   & 0.5433     & 0.6266     & 0.6806    \\
w/o Reward-PR & 0.0694    & 0.3987   & 0.5923   & 0.4502     & 0.5335     & 0.6087    \\
w/o MCTS-TTS  & 0.1307    & 0.4886   & 0.6462   & 0.5082     & 0.6185     & 0.6511    \\ 
\hline
\multicolumn{7}{c}{Round 3}                                                           \\ \hline
StageMCTS     & 0.1462    & 0.5221   & 0.6552   & 0.5392     & 0.6266     & 0.6838    \\
w/o Reward-PR & 0.0629    & 0.3930   & 0.5792   & 0.4526     & 0.5368     & 0.6176    \\
w/o MCTS-TTS  & 0.1413    & 0.5123   & 0.6479   & 0.5180     & 0.6225     & 0.6601    \\ 
\hline
\end{tabular}
}
\end{table}

%% file: tables/case_study215.tex
\begin{table*}[htbp]
\centering
\caption{Case study of queries and corresponding results from different methods for a question selected from the test set. \correct{Green} indicates a \correct{correct} answer that matches the ground-truth. \wrong{Red} indicates a \wrong{wrong} answer with incorrect results.}
\label{tab:case_study_215}
\begin{tabularx}{\linewidth}{@{}l>{\raggedright\arraybackslash}X@{}}
\toprule
\textbf{Method} & \textbf{Query Statement and Result} \\
\midrule
\textbf{Dialogue History} & $2$ rounds omitted \\
\midrule
\textbf{Question (ID: $215$)} & How many professionals have at most 3 treatments? \\
\midrule
\textbf{Ground-Truth} & 
\compactcode{db.Professionals.aggregate([ \{ \$unwind: "\$Treatments" \}, \{ \$group: \{ \_id: "\$professional\_id", treatmentCount: \{ \$sum: 1 \} \} \}, \{ \$match: \{ treatmentCount: \{ \$lte: 3 \} \} \}, \{ \$group: \{ \_id: null, count: \{ \$sum: 1 \} \} \}, \{ \$project: \{ \_id: 0, count: 1 \} \} ])}
\newline
\textbf{Result:} \correct{\texttt{[{'count': 8}]}} \\
\cmidrule{1-2}
\textbf{Stage-MCTS (Ours)} & 
\compactcode{db.Professionals.aggregate([ \{ \$unwind: "\$Treatments" \}, \{ \$group: \{ \_id: "\$professional\_id", treatCount: \{ \$sum: 1 \} \} \}, \{ \$match: \{ treatCount: \{ \$lte: 3 \} \} \}, \{ \$count: "profsWithFewTreatments" \} ]);}
\newline
\textbf{Result:} \correct{\texttt{[{'profsWithFewTreatments': 8}]}} \\
\cmidrule{1-2}
\textbf{Zero-Shot (GPT-4o)} & 
\compactcode{db.Professionals.aggregate([ \{ \$addFields: \{ treatmentCount: \{ \$size: "\$Treatments" \} \} \}, \{ \$match: \{ treatmentCount: \{ \$lte: 3 \} \} \}, \{ \$count: "professionalsWithAtMost3Treatments" \} ]);}
\newline
\textbf{Result:} \wrong{\texttt{[{'professionalsWithAtMost3Treatments': 15}]}} \\
\cmidrule{1-2}
\textbf{STaR-SFT (DeepSeek-R1 Data)} & 
\compactcode{db.Professionals.aggregate([ \{ \$addFields: \{ treatmentCount: \{ \$size: "\$Treatments" \} \} \}, \{ \$match: \{ treatmentCount: \{ \$lte: 3 \} \} \}, \{ \$count: "number\_of\_professionals" \} ])}
\newline
\textbf{Result:} \wrong{\texttt{[{'number\_of\_professionals': 15}]}} \\
\bottomrule
\end{tabularx}
\end{table*}

%% file: tables/ablation_study_reward_model.tex
\begin{table}[htbp]
\centering
\caption{Ablation study on the test-time scaling strategies, based on the SLM after three rounds of self-training.}
\label{tab:ablation_study_reward_model}
\scalebox{.82}{
\begin{tabular}{lllllll}
\hline
           & \multicolumn{3}{c}{Query-based} & \multicolumn{3}{c}{Execution-based} \\ \hline
Method     & EM        & QSM      & QFC      & EX         & EFM        & EVM       \\ \hline
No TTS     & 0.1413    & 0.5123   & 0.6479   & 0.5180     & 0.6225     & 0.6601    \\ \hline
Self-consistency  & \textbf{0.1503} & \textbf{0.5319} & 0.6528          & 0.5319          & 0.6111          & 0.6650          \\
BoN + PRM  & 0.1291    & 0.4951   & 0.6283   & 0.4935     & 0.5817     & 0.6348    \\
MCTS + PRM & 0.1405    & 0.5245   & 0.6487   & 0.5180     & 0.6119     & 0.6495    \\
MCTS + SRM (ours) & 0.1462          & 0.5221          & \textbf{0.6552} & \textbf{0.5392} & \textbf{0.6266} & \textbf{0.6838} \\ \hline
\end{tabular}
}
\end{table}

%% file: tables/case_study182.tex
\begin{table*}[htbp]
\centering
\caption{Case study of queries and corresponding results from different methods for a question selected from the test set. \correct{Green} indicates a \correct{correct} answer that matches the ground-truth. \wrong{Red} indicates a \wrong{wrong} answer with execution errors or incorrect results. \partialcorrect{Orange} indicates a \partialcorrect{partially correct} answer with some issues.}
\label{tab:case_study_182}
\begin{tabularx}{\linewidth}{@{}l>{\raggedright\arraybackslash}X@{}}
\toprule
\textbf{Method} & \textbf{Query Statement and Result} \\
\midrule
\textbf{Dialogue History} & 5 rounds omitted \\
\midrule
\textbf{Question (ID: $182$)} & For the cars with 4 cylinders, which model has the largest horsepower? \\
\midrule
\textbf{Ground-Truth} & 
\compactcode{db.continents.aggregate([\{ \$unwind: "\$countries" \}, \{ \$unwind: "\$countries.car\_makers" \}, \{ \$unwind: "\$countries.car\_makers.model\_list" \}, \{ \$unwind: "\$countries.car\_makers.model\_list.car\_names" \}, \{ \$unwind: "\$countries.car\_makers.model\_list.car\_names.cars\_data" \}, \{ \$match: \{ \$expr: \{ \$and: [ \{ \$eq: ["\$countries.car\_makers.model\_list.car\_names.cars\_data.Cylinders", 4] \}, \{ \$eq: ["\$countries.car\_makers.model\_list.car\_names.MakeId", "\$countries.car\_makers.model\_list.car\_names.cars\_data.Id"] \} ] \} \} \}, \{ \$sort: \{ "countries.car\_makers.model\_list.car\_names.cars\_data.Horsepower": -1 \} \}, \{ \$limit: 1 \}, \{ \$project: \{ \_id: 0, Model: "\$countries.car\_makers.model\_list.car\_names.Model" \} \}])}
\newline
\textbf{Result:} \correct{\texttt{[{'Model': 'amc'}]}} \\
\cmidrule{1-2}
\textbf{Stage-MCTS (Ours)} & 
\compactcode{db.continents.aggregate([\{ \$unwind: "\$countries" \}, \{ \$unwind: "\$countries.car\_makers" \}, \{ \$unwind: "\$countries.car\_makers.model\_list" \}, \{ \$unwind: "\$countries.car\_makers.model\_list.car\_names" \}, \{ \$unwind: "\$countries.car\_makers.model\_list.car\_names.cars\_data" \}, \{ \$match: \{ "countries.car\_makers.model\_list.car\_names.cars\_data.Cylinders": 4 \} \}, \{ \$sort: \{ "countries.car\_makers.model\_list.car\_names.cars\_data.Horsepower": -1 \} \}, \{ \$limit: 1 \}, \{ \$project: \{ \_id: 0, Model: "\$countries.car\_makers.model\_list.car\_names.Model" \} \}]);}
\newline
\textbf{Result:} \correct{\texttt{[{'Model': 'amc'}]}} \\
\cmidrule{1-2}
\textbf{Zero-Shot (GPT-4o)} & 
\compactcode{db.continents.aggregate([\{ \$unwind: "\$countries" \}, \{ \$unwind: "\$countries.car\_makers" \}, \{ \$unwind: "\$countries.car\_makers.model\_list" \}, \{ \$unwind: "\$countries.car\_makers.model\_list.car\_names" \}, \{ \$unwind: "\$countries.car\_makers.model\_list.car\_names.cars\_data" \}, \{ \$match: \{ "countries.car\_makers.model\_list.car\_names.cars\_data.Cylinders": 4 \} \}, \{ \$project: \{ "Model": "\$countries.car\_makers.model\_list.car\_names.Model", "Horsepower": \{ \$toInt: "\$countries.car\_makers.model\_list.car\_names.cars\_data.Horsepower" \}, "\_id": 0 \}\}, \{ \$sort: \{ "Horsepower": -1 \} \}, \{ \$limit: 1 \}])}
\newline
\textbf{Result:} \wrong{\texttt{Query execution error: QUERY\_ERROR: PlanExecutor error during aggregation :: caused by :: Failed to parse number 'null' in \$convert with no onError value: Did not consume whole string.}} \\
\cmidrule{1-2}
\textbf{STaR-SFT (DeepSeek-R1 Data)} & 
\compactcode{db.continents.aggregate([\{ \$unwind: "\$countries" \}, \{ \$unwind: "\$countries.car\_makers" \}, \{ \$unwind: "\$countries.car\_makers.model\_list" \}, \{ \$unwind: "\$countries.car\_makers.model\_list.car\_names" \}, \{ \$unwind: "\$countries.car\_makers.model\_list.car\_names.cars\_data" \}, \{ \$match: \{ "countries.car\_makers.model\_list.car\_names.cars\_data.Cylinders": 4 \} \}, \{ \$project: \{ model: "\$countries.car\_makers.model\_list.car\_names.Model", horsepower: "\$countries.car\_makers.model\_list.car\_names.cars\_data.Horsepower" \} \}, \{ \$group: \{ \_id: null, maxHorsepower: \{ \$max: "\$horsepower" \}, allDocs: \{ \$push: "\$\$ROOT" \} \} \}, \{ \$unwind: "\$allDocs" \}, \{ \$match: \{ \$expr: \{ \$eq: ["\$allDocs.horsepower", "\$maxHorsepower"] \} \} \}, \{ \$project: \{ \_id: 0, Model: "\$allDocs.model" \} \}, \{ \$group: \{ \_id: "\$Model", Model: \{ \$first: "\$Model" \} \} \}, \{ \$project: \{ \_id: 0, Model: 1 \} \}])}
\newline
\textbf{Result:} \partialcorrect{\texttt{[{'Model': 'renault'}, {'Model': 'amc'}, {'Model': 'ford'}]}} \\
\bottomrule
\end{tabularx}
\end{table*}

%% file: appendices/prompts.tex
\section{Prompt Template}
The prompt templates used in this study are detailed below.

\begin{figure*}[htbp]  
\centering
\begin{lstlisting}[breaklines=true]
You are a MongoDB aggregation expert. Your task is to analyze a given final solution and reconstruct the thinking process that led to it.

**WORKFLOW:**
1. **Analyze the Final Solution:** Understand the complete aggregation pipeline
2. **Deconstruct the Pipeline:** Break down the pipeline into logical steps
3. **Step-by-Step Reasoning:** For each step, explain the purpose and logic
4. **Connect Steps:** Show how each step builds on previous ones
5. **Justify Design Choices:** Explain why specific operators or approaches were used

**OUTPUT FORMAT (strictly follow):**

<think>
<step>Step 1: [Analyze what this stage accomplishes and why it's positioned here]</step>
<draft>[Copy the MongoDB aggregation stage(s) for this step]</draft>
<reasoning>[Explain the purpose, logic, and any performance considerations]</reasoning>

<step>Step 2: [Analyze what this stage accomplishes and why it's positioned here]</step>
<draft>[Copy the MongoDB aggregation stage(s) for this step]</draft>
<reasoning>[Explain the purpose, logic, and how it builds on previous steps]</reasoning>

...
<step>Step N: [Analyze what this stage accomplishes and why it's positioned here]</step>
<draft>[Copy the MongoDB aggregation stage(s) for this step]</draft>
<reasoning>[Explain the purpose, logic, and final shaping of results]</reasoning>
</think>

**REQUIREMENTS:**
- Deconstruct the given final solution into logical steps
- Each step should include: analysis, draft stage(s), and reasoning
- Explain the strategic purpose of each stage (why it's needed, why it's placed there)
- Mention any performance optimizations or design choices
- Connect steps to show progressive data transformation
- Do NOT generate new solutions - only analyze and explain the given one

## CONTEXT INFORMATION

**DATABASE SCHEMA:**
{SCHEMA_CONTEXT}

**DIALOGUE HISTORY:**
{DIALOGUE_HISTORY}
(Note: "No history" indicates first interaction.)

**ORIGINAL QUESTION:**
{QUESTION}

**FINAL SOLUTION (Target Query):**
{FINAL_SOLUTION}

**YOUR TASK:**
Analyze the given final solution and generate ONLY the thinking process that would lead to this solution. Your output should:

1. Deconstruct the aggregation pipeline into logical steps
2. For each step, explain:
   - What the step accomplishes
   - Why it's positioned at that point in the pipeline
   - How it transforms the data
   - Any performance or design considerations
3. Show the progressive flow from one step to the next
4. Do NOT modify the solution or suggest alternatives

**OUTPUT FORMAT REQUIREMENTS:**
- Only output the <think> section with steps
- Each step must have three parts: <step>, <draft>, and <reasoning>
- Do NOT include <answer> section or final pipeline (it's already provided)
- Ensure explanations reference the schema fields appropriately
\end{lstlisting}
\caption{Prompt for Path Refinement.}
\label{prompt:path_refinement}
\end{figure*}

\begin{figure*}[htbp]  
\centering
\begin{lstlisting}[breaklines=true]
<GENERATE-ONE-STEP>
You are a MongoDB aggregation expert. Generate queries step-by-step following these rules:

**WORKFLOW:**
1. **If no solution exists:** Start from scratch - provide the first logical step
2. **If solution is incomplete:** Provide ONLY the next logical step
3. **If solution is complete:** Provide ONLY the final query

**OUTPUT FORMAT (strictly follow one):**

1. **For intermediate steps:**
<step>Step [number]: [Brief reasoning about this stage's purpose]</step>
<draft>[draft MongoDB aggregation stage(s)]</draft>

Example:
<step>Step 1: Filter active users by location using $match</step>
<draft>{ $match: { status: "active", location: "NYC" } }</draft>

2. **For final answer:**
<answer>Final Answer: ```mql
[Complete aggregation pipeline]
```</answer>
Example: `<answer>Final Answer: ```mql\ndb.department.aggregate([\n  { $unwind: "$instructors" }\n]);\n```</answer>`

**BEST PRACTICES:**
- Filter early with `$match` to reduce data
- Use `$lookup` only when necessary joins are required
- Handle arrays with `$unwind` or `$group` appropriately
- Shape output with `$project` or `$addFields` at appropriate stages
- Consider performance implications of each stage

## CONTEXT

**DATABASE SCHEMA:**
{SCHEMA_CONTEXT}

**DIALOGUE HISTORY:**
{DIALOGUE_HISTORY}
(Note: Turns separated by "|". "No history" indicates first interaction.)

**CURRENT TASK**
Question: {QUESTION}

**CURRENT PROGRESS:**
{EXISTING_SOLUTION}

**HINTS:**
{HINT}

**ACTION REQUIRED:**
Analyze the current state and provide:
- The next aggregation step(s) with reasoning, OR
- The final complete query

**IMPORTANT:** Output ONLY in the specified format above.
\end{lstlisting}
\caption{Prompt for next-step generation.}
\label{prompt:generate_next_step}
\end{figure*}

\begin{figure*}[htbp]  
\centering
\begin{lstlisting}[breaklines=true]
<ROLLOUT>
You are a MongoDB aggregation expert. Generate the complete query with the remaining steps explained in one response, given the existing solution.

**WORKFLOW:**
1. **Analyze the problem:** Understand the requirements based on schema, question, and existing solution
2. **Plan all steps:** Design the complete aggregation pipeline with logical reasoning for each stage
3. **Provide complete solution:** Output all steps followed by the final query

**OUTPUT FORMAT (strictly follow):**

1. **First, output the remaining steps with reasoning:**
<step>Step i+1: [Brief reasoning about this stage's purpose]</step>
<draft>[draft MongoDB aggregation stage(s)]</draft>

<step>Step i+2: [Brief reasoning about this stage's purpose]</step>
<draft>[draft MongoDB aggregation stage(s)]</draft>

<step>Step i+N: [Brief reasoning about this stage's purpose]</step>
<draft>[draft MongoDB aggregation stage(s)]</draft>

2. **Then, output the complete final answer:**
<answer>Final Answer: ```mql
[Complete aggregation pipeline]
```</answer>

**Example output format:** (For the existing solution)
{EXAMPLE}

**BEST PRACTICES:**
- Filter early with `$match` to reduce data
- Use `$lookup` only when necessary joins are required
- Handle arrays with `$unwind` or `$group` appropriately
- Shape output with `$project` or `$addFields` at appropriate stages
- Consider performance implications of each stage
- Keep the pipeline logical and efficient

## CONTEXT

**DATABASE SCHEMA:**
{SCHEMA_CONTEXT}

**DIALOGUE HISTORY:**
{DIALOGUE_HISTORY}
(Note: Turns separated by "|". "No history" indicates first interaction.)

**CURRENT TASK**
Question: {QUESTION}

**CURRENT PROGRESS:**
{EXISTING_SOLUTION}

**HINTS:**
{HINT}

**ACTION REQUIRED:**
Analyze the current state and provide:
1. All intermediate aggregation steps with reasoning for each stage
2. The final complete aggregation pipeline

**IMPORTANT:** 
- Output ALL steps in the <step> section with <draft>
- Output the complete final query in the <answer> section
- If existing solution is complete, just output it in <answer> with a brief note
- Follow the exact output format specified above
\end{lstlisting}
\caption{Prompt for rollout.}
\label{prompt:rollout}
\end{figure*}

\begin{figure*}[htbp]  
\centering
\begin{lstlisting}[breaklines=true]
<step>Step i+1: Filter active users by location using $match</step>
<draft>{ $match: { status: "active", location: "NYC" } }</draft>

<step>Step i+2: Join with department collection using $lookup</step>
<draft>{ $lookup: { from: "departments", localField: "dept_id", foreignField: "_id", as: "department" } }</draft>

<step>Step i+3: Flatten department array using $unwind</step>
<draft>{ $unwind: "$department" }</draft>

<step>Step i+4: Group users by department and calculate average salary</step>
<draft>{ $group: { _id: "$department.name", avgSalary: { $avg: "$salary" } } }</draft>

<step>Step i+5: Sort results by average salary descending</step>
<draft>{ $sort: { avgSalary: -1 } }</draft>

<answer>Final Answer: ```mql
db.users.aggregate([
  { $match: { status: "active", location: "NYC" } },
  { $lookup: { from: "departments", localField: "dept_id", foreignField: "_id", as: "department" } },
  { $unwind: "$department" },
  { $group: { _id: "$department.name", avgSalary: { $avg: "$salary" } } },
  { $sort: { avgSalary: -1 } }
]);
```</answer>
\end{lstlisting}
\caption{An example employed in the rollout prompt.}
\label{prompt:example_in_rollout}
\end{figure*}

\begin{figure*}[htbp]  
\centering
\begin{lstlisting}[breaklines=true]
You are a translation system designed to convert SQL queries into MongoDB queries. You will be provided with details about a MongoDB collection, a user's natural language description of their data retrieval needs, and the SQL query that represents this need. Your task is to analyze the SQL query and translate it into an equivalent MongoDB query.

Here are the details you will receive:

MongoDB Collection and its Fields: {COLLECTION_INFO}

User Natural Language Question: {QUESTION}

Corresponding SQL Query: {SQL_QUERY}

Let's think step by step. Using the information provided, carefully convert the SQL query into a MongoDB query that retrieves the same data. Consider all aspects of the SQL query, such as selection criteria, sorting, and joining conditions, and reflect these in your MongoDB query translation. 

Provide your answer in the format as follows:

Explanation: (Explanation for the MongoDB Query)
Answer: 
###
(MongoDB Query)
###

{DIALOGUE_EXAMPLES}
\end{lstlisting}
\caption{Prompt for dataset re-generation.}
\label{prompt:dataset_regeneration}
\end{figure*}

\begin{figure*}[htbp]  
\centering
\begin{lstlisting}[breaklines=true]
You are a helpful assistant that can answer questions and help with tasks, especially for tasks that are related to the MongoDB database.

You are given a natural language query, a database schema, and a list of previous interactions.
You need to answer the natural language query based on the database schema and the previous interactions.
Specifically, the history interactions are the dialogue history between the user and the assistant. Each turn is separated by a special token "|". If there is no history, it will be "No history provided. This is the first turn."

Natural Language Query: {QUESTION}
Database Schema: {SCHEMA_CONTEXT}
Previous Interactions: {DIALOGUE_HISTORY}

You should only output the MQL query, no other text. You'd better be sure about the answer.
The format of the MQL query is:
```mongodb
db.collection.aggregate();
```
or
```mongodb
db.collection.find();
```

{EXAMPLES}
\end{lstlisting}
\caption{Prompt for the training-free paradigms.}
\label{prompt:in_context_learning}
\end{figure*}

\begin{figure*}[htbp]  
\centering
\begin{lstlisting}[breaklines=true]
You are an experienced database expert.
Now you need to generate a MongoDB query given the database information, a natural language query and some additional information.
The database structure is defined by the following database schema (a dictionary of collections, their fields and their types).
The dialogue history is the dialogue history between the user and the assistant. Each turn is separated by a special token "|". If there is no history, it will be "No history provided. This is the first turn."

Given the database schema,the `Natural Language Query` and the `Dialogue History`. You will be given database schema and you need understand the database and fields.

You will be using a way called "recursive divide-and-conquer approach to MongoDB query generation from natural language".

Here is a high level description of the steps.
1. **Divide (Decompose Sub-question with Pseudo MongoDB):** The complex natural language query is recursively broken down into simpler sub-questions. Each sub-question targets a specific piece of information or logic required for the final MongoDB query. 
2. **Conquer (Real MongoDB for sub-questions):**  For each sub-question (and the main question initially), a "pseudo-MongoDB" fragment is formulated. This pseudo-MongoDB represents the intended MongoDB logic but might have placeholders for answers to the decomposed sub-questions. 
3. **Combine (Reassemble):** Once all sub-questions are resolved and their corresponding MongoDB fragments are generated, the process reverses. The MongoDB fragments are recursively combined by replacing the placeholders in the pseudo-MongoDB with the actual generated MongoDB from the lower levels.
4. **Final Output:** This bottom-up assembly culminates in the complete and correct MongoDB query that answers the original complex question.

Database admin instructions (violating any of the following will result in punishable to death!):
{ADMIN_INSTRUCTION}

Repeating the question and hint, and generating the MongoDB query with Recursive Divide-and-Conquer.

Please respond with xml tags structured as follows:
```xml
<think>
Your detailed reasoning for the MongoDB query generation, with Recursive Divide-and-Conquer approach.
</think>
<mql>
The final MongoDB query that answers the question. The format of the MongoDB query is:
db.collection.aggregate();
or
db.collection.find();
</mql>
```

**DATABASE SCHEMA:**
{SCHEMA_CONTEXT}

**DIALOGUE HISTORY:**
{DIALOGUE_HISTORY}
(Note: Turns separated by "|". "No history" indicates first interaction.)

**CURRENT TASK**
Question: {QUESTION}

**HINTS:**
{HINT}

Only output xml format (starting with ```xml and ending with ```) as your response.
\end{lstlisting}
\caption{Prompt for STaR-SFT.}
\label{prompt:star_sft}
\end{figure*}

\begin{figure*}[htbp]  
\centering
\begin{lstlisting}[breaklines=true]
1. **Projection:**
    - Only project fields mentioned in the user's question using `$project`.
    - Avoid unnecessary fields or values.
2. **Aggregation Operations:**
    - Always perform `$lookup` before using `$max` or `$min`.
    - Use `$group` with accumulators like `$max`, `$min`, `$sum`, `$avg`.
3. **Sorting with Distinct Values:**
    - Use `$group` before `$sort` to ensure distinct values.
4. **Handling Null/Missing Fields:**
    - Use `$exists` or `$ne: null` in `$match` stage to handle null/missing fields.
    - Consider using `$lookup` with proper matching conditions.
5. **Collection References:**
    - Only include collections essential to answer the question.
6. **Strictly Follow Hints:**
    - Adhere to all provided hints.
7. **Thorough Question Analysis:**
    - Address all conditions mentioned in the question.
8. **Distinct Values:**
    - Use `$group` when unique values are required (e.g., IDs, URLs).
9. **Field Selection:**
    - Carefully analyze field descriptions and hints to choose correct fields when similar fields exist across collections.
10. **String Operations:**
    - Use `$concat` for string concatenation only when absolutely necessary.
11. **MongoDB Operators Only:**
    - Use only operators available in MongoDB.
12. **Date Processing:**
    - Use `$dateToString` for date formatting
    - Use date operators like `$year`, `$month`, `$dayOfMonth` for date parts.
13. **Field References:**
    - Use dot notation for nested fields and array elements.
    - Reference fields as `$fieldName` in expressions.
14. **Join Operations:**
    - Prioritize `$lookup` with `$unwind` over nested queries.
    - Use `$lookup` with pipeline for complex joins.
\end{lstlisting}
\caption{Database Administrator instructions used in STaR-SFT.}
\label{prompt:admin_instruction}
\end{figure*}

\begin{figure*}[htbp]  
\centering
\begin{lstlisting}[breaklines=true]
You are a helpful assistant in a MongoDB database.
You are given a database schema, the dialogue history, and the natural language query.
The database schema is a dictionary containing the collection names, their fields and their types.
The dialogue history is a list of tuples, each tuple contains a question and a MongoDB query. Each turn is separated by a special token "|". If there is no history, it will be "No history provided. This is the first turn."
The natural language query is the question to answer.
The objective is to decompose the natural language query into a plan of steps to answer the question. The first step should decide which collection to use and which methods like find or aggregate to use.
Each step should be corresponding to a stage in the aggregation pipeline of the MongoDB query. It should be in English instead of code, like "Group the documents by 'Directed_by' field and sum up the 'Show_times_per_day' from the unwound documents to calculate the total show times for each director."

Wrap the output in `json` format.

The database schema is:
{db_schema}
The dialogue history is:
{dialogue_history}
The natural language query is:
{natural_language_query}
\end{lstlisting}
\caption{Prompt for the ReAct agent.}
\label{prompt:react}
\end{figure*}

\begin{figure*}[htbp]  
\centering
\begin{lstlisting}[breaklines=true]
For the given objective, come up with a simple step by step plan and the corresponding JSON format MongoDB query clause. 
This plan should involve individual tasks, such as "Use the MongoDB aggregation framework to unwind the nested arrays". Do not add any superfluous steps. Since you could not connect to the database and access the data, you can analyze some plans to get the information you need. You should make full use of the MongoDB schema given to you.
Make sure that each step has all the information needed - do not skip steps.

For the given objective, the current plan and the corresponding JSON format MongoDB query clause, please generate the valid MongoDB query. \
The plan involves individual steps, such as "Use the MongoDB aggregation framework to unwind the nested arrays.". \
The corresponding JSON format MongoDB query clause involves the aggregation framework, such as "$unwind". \
The response should be a valid MongoDB query only. For example, db.collection_name.find(<query>) or db.collection_name.aggregate(<pipeline>).

Your objective was this:
{input}

Your plan was this:
{plan}

The corresponding JSON format MongoDB query clause was this:
{solution}

Please output the final response in the following format:
db.collection_name.find(<query>)
db.collection_name.aggregate(<pipeline>)

Let's think step by step.
\end{lstlisting}
\caption{Prompt for the Plan-and-Solve agent.}
\label{prompt:plan_and_solve}
\end{figure*}